\documentclass[floats,aps,amsmath,showpacs,nofootinbib]{revtex4}

\usepackage{graphicx}

\usepackage{hhline} \newcommand{\tr}{{\rm tr}}
 
\newcommand{\ignore}[1]{} \newcommand{\vv}{\vec{v}}
\newcommand{\p}{{\cal P}}

\font\eightln=line10 at8pt \catcode`\@=11
\def\singlearrow{\@ifnextchar [{\@singlearrow }{\@singlearrow[0]}}
\def\@singlearrow[#1]{\mathrel{\,\lower0.15ex
    \hbox{\let\@linefnt\eightln\unitlength0.6ex\begin{picture}(4,3)
        \put(0,1.5){\vector(1,0){4}}
    \end{picture}}\,}}
\catcode`\@=12

\font\eightln=line10 at8pt \catcode`\@=11
\def\doublearrow{\@ifnextchar [{\@doublearrow }{\@doublearrow[0]}}
\def\@doublearrow[#1]{\mathrel{\,\lower0.15ex
    \hbox{\let\@linefnt\eightln\unitlength0.6ex\begin{picture}(4,3)
        \ifcase#1\put(4,0.8){\vector(-1,0){4}}\put(0,2){\vector(1,0){4}}
        \or \put(0,0.8){\vector(1,0){4}}\put(0,2){\vector(1,0){4}}\fi
    \end{picture}}\,}}
\catcode`\@=12

\font\eightln=line10 at8pt \catcode`\@=11
\def\triplearrow{\@ifnextchar [{\@triplearrow }{\@triplearrow[0]}}
\def\@triplearrow[#1]{\mathrel{\,\lower0.15ex
    \hbox{\let\@linefnt\eightln\unitlength0.6ex\begin{picture}(4,3)
        \ifcase#1\put(0,0.3){\vector(1,0){4}}\put(0,1.5){\vector(1,0){4}}
        \put(0,2.7){\vector(1,0){4}}
        \or\put(0,0){\vector(2,1){4}}\put(0,1){\vector(2,1){4}}
        \put(0,3){\vector(4,-3){4}}
        \or\put(0,3){\vector(2,-1){4}}\put(0,2){\vector(2,-1){4}}
        \put(0,0){\vector(4,3){4}}
        \or\put(0,0){\vector(4,3){4}}\put(0,1.5){\vector(1,0){4.5}}
        \put(0,3){\vector(4,-3){4}}
        \or\put(0,0.3){\vector(1,0){4}}\put(0,1){\vector(2,1){4}}
        \put(0,3){\vector(2,-1){4}}
        \or\put(0,0){\vector(2,1){4}}\put(0,2){\vector(2,-1){4}}
        \put(0,2.7){\vector(1,0){4}}
        \or\put(4,0.3){\vector(-1,0){4}}\put(4,1.5){\vector(-1,0){4}}
        \put(0,2.7){\vector(1,0){4}}
        \or\put(4,2){\vector(-2,-1){4}}\put(4,3){\vector(-2,-1){4}}
        \put(0,3){\vector(4,-3){4}}
        \or\put(4,0){\vector(-1,0){4}}\put(4,3){\vector(-2,-1){4}}
        \put(0,3){\vector(2,-1){4}}
        \or\put(4,0.3){\vector(-1,0){4}}\put(0,1.5){\vector(1,0){4}}
        \put(0,2.7){\vector(1,0){4}} \fi
  \end{picture}}\,}}
\catcode`\@=12

\newcommand{\asr}{\doublearrow} \newcommand{\psr}{\doublearrow[1]}
\newcommand{\pc}{\triplearrow} \newcommand{\ac}{\triplearrow[9]}

\makeatletter
\newcommand{\contraction}[5][1ex]{%
  \mathchoice
    {\contraction@\displaystyle{#2}{#3}{#4}{#5}{#1}}%
    {\contraction@\textstyle{#2}{#3}{#4}{#5}{#1}}%
    {\contraction@\scriptstyle{#2}{#3}{#4}{#5}{#1}}%
    {\contraction@\scriptscriptstyle{#2}{#3}{#4}{#5}{#1}}}%
\newcommand{\contraction@}[6]{%
  \setbox0=\hbox{$#1#2$}%
  \setbox2=\hbox{$#1#3$}%
  \setbox4=\hbox{$#1#4$}%
  \setbox6=\hbox{$#1#5$}%
  \dimen0=\wd2%
  \advance\dimen0 by \wd6%
  \divide\dimen0 by 2%
  \advance\dimen0 by \wd4%
  \vbox{%
    \hbox to 0pt{%
      \kern \wd0%
      \kern 0.5\wd2%
      \contraction@@{\dimen0}{#6}%
      \hss}%
    \vskip 0.2ex%
    \vskip\ht2}}

\newcommand{\contraction@@}[3][0.06em]{%
  \hbox{%
    \vrule width #1 height 0pt depth #3%
    \vrule width #2 height 0pt depth #1%
    \vrule width #1 height 0pt depth #3%
    \relax}}

\begin{document}

\newcommand{\QATOP}[2]{#1\atop #2}

\title{Semiclassical Approach to Chaotic Quantum Transport}

\date{\today }
\author{ Sebastian M{\"u}ller$^1$, Stefan Heusler$^2$,
 Petr Braun$^{2,3}$, Fritz Haake$^2$}

\address{
$^1$Cavendish Laboratory, University of Cambridge, J J Thomson Avenue,
Cambridge CB3 0HE, UK\\
$^2$Fachbereich Physik, Universit{\"a}t Duisburg-Essen,
47048 Duisburg, Germany\\
$^3$Institute of Physics, Saint-Petersburg University, 198504
Saint-Petersburg, Russia}

\begin{abstract}
  We describe a semiclassical method to calculate universal transport
  properties of chaotic cavities. While the energy-averaged
  conductance turns out governed by pairs of entrance-to-exit
  trajectories, the conductance variance, shot noise and other related
  quantities require trajectory quadruplets; simple diagrammatic
  rules allow to find the contributions of these pairs and quadruplets.
  Both pure symmetry classes and the crossover due to an external
  magnetic field are considered.
\end{abstract}
 \maketitle

\ignore{
\tableofcontents
\newpage}

\section{INTRODUCTION}

Mesoscopic cavities show universal transport properties -- such as 
conductance, conductance fluctuations, or shot noise --
provided the classical dynamics inside the cavity is fully chaotic.
Here chaos may be due to either implanted impurities or bumpy boundaries.
A phenomenological description of these universal features
is available through random-matrix theory
(RMT) by averaging over {\it ensembles} of systems (whose Hamiltonians
are represented by matrices) \cite{Beenakker}. For systems with impurities,
one can alternatively average over different disorder potentials.
However, experiments show that even {\it individual} cavities show universal
behavior faithful to these averages.

In the present paper we want to show why this is the case. To do so
we propose a semiclassical explanation of universal transport through
individual chaotic cavities, based on the interfering contributions of
close classical trajectories.  This approach generalizes earlier work
in \cite{RS,Schanz,EssenCond,EssenShot} and is inspired by recent
progress for universal spectral statistics
\cite{Berry,Argaman,SR,EssenFF}. Our semiclassical procedure  often
turns out to be technically easier than RMT; transport properties are
evaluated through very simple diagrammatic rules.

We consider a  two-dimensional cavity accommodating chaotic
classical motion. Two (or more) straight leads are attached to the
cavity and carry currents. We shall mostly consider electronic
currents, but most of the following ideas apply to transport of
light or sound as well, minor modifications apart.

The leads support wave modes (``channels'') ${\rm e}^{{{\rm i}} k
  x_i\cos\theta_i}\sin(ky_i\sin|\theta_i|)$; the subscripts $i=1,2$
refer to the ingoing and outgoing lead, respectively, $x_i$ and
$y_i$ with $0<y_i<w_i$ are coordinates along and transversal to
the lead.
Here, $w_i$ is the width of the lead, $k$ the wave number, and
$\theta_i$ the angle enclosed between the wave vector and the
direction of the lead.
Dirichlet boundary conditions inside the
lead impose the restriction $kw_i\sin|\theta_i|=a_i\pi$ with the
channel index $a_i$ running from 1 to $N_i$, the largest integer
below $\frac{kw_i}{\pi}$. Classically, the $a_i$-th channel can be
associated with trajectories inside the lead that enclose an angle
$\theta_i$ with the lead direction,
regardless of their location in configuration space. 
The sign of the enclosed angle
changes after each reflection at the boundaries of the cavity, and
angles of both signs are associated to the same channel.

We shall determine, e.g., the mean and the variance of the conductance
as power series in the inverse of the number of channels $N=N_1+N_2$.
In contrast to much of the previous literature, we will go to all
orders in $\frac{1}{N}$.  We shall be interested both in dynamics with
time reversal invariance (``orthogonal case'') and without that
symmetry (``unitary case'').  For electronic motion time reversal
invariance may be broken by an external magnetic field. For that
latter case, we shall also interpolate between both pure symmetry
classes by account for a weak magnetic field producing magnetic
actions of the order of $\hbar$.

We will always work in the semiclassical limit, and thus require the
linear dimension $L$ of the cavity to be large compared to the
(Fermi)
wavelength $\lambda$. When taking the limit $\frac{\lambda}{L}\to 0$,
the number of channels $N\propto\frac{w}{\lambda}$ ($w\sim w_1\sim
w_2$) will be increased  only slowly. The width of the
openings thus becomes small compared to $L$. For this particular
semiclassical limit, the dwell time of trajectories inside the cavity,
$T_D\propto \frac{L}{w}$ grows faster than the so-called
Ehrenfest time $T_E\propto \ln\frac{L}{\lambda}\propto\ln\hbar$.
Interesting effects arising for $T_E/T_D$ of order unity
\cite{Brouwer,Whitney,EhrenfestRMT} are thus discarded.

Following Landauer and B{\"u}ttiker \cite{Landauer,Buettiker}, we view
transport as scattering between leads and deal with amplitudes
$t_{a_1a_2}$ for transitions between channels $a_1$ and $a_2$. These
amplitudes form an $N_1\times N_2$ matrix $t=\{t_{a_1a_2}\}$. Each
$t_{a_1a_2}$ can be approximated semiclassically, by the van Vleck
approximation for the propagator, as a sum over trajectories
connecting the channels $a_1$ and $a_2$ \cite{Richter},
\begin{equation}
\label{transition}
t_{a_1a_2}\sim\frac{1}{\sqrt{T_H}}\sum_{\alpha:a_1\to a_2}
A_\alpha {\rm e}^{{{\rm i}} S_\alpha/\hbar} \,.
\end{equation}
The channels exactly determine the absolute values of the initial
and final angles of incidence $\theta_1,\theta_2$ of the
contributing trajectories; again both positive and negative angles
are possible.
In (\ref{transition}), $T_H$ denotes the so-called
Heisenberg time, i.e., the quantum time scale
$2\pi\hbar\overline{\rho}$ associated to the mean level density
$\overline{\rho}$.
The Heisenberg time diverges in the
semiclassical limit like
$T_H=2\pi\hbar\overline{\rho}\sim\frac{\Omega}{(2\pi\hbar)^{f-1}}$
with $\Omega$ the volume of the energy shell and $f$ the number of
degrees of freedom; we shall mostly consider $f=2$. The
``stability amplitude'' $A_\alpha$  (which includes the so-called Maslov index)
can be found in Richter's
review \cite{Richter}. Finally, the phase in (\ref{transition})
depends on the classical action $S_\alpha=\int_\alpha{\bf p}\cdot
d{\bf q}$.

Within the framework just delineated, we will evaluate, for individual
fully chaotic cavities,
\begin{itemize}
\item the {\it mean conductance}
  $\big\langle\tr(tt^\dagger)\big\rangle$ (Actually, the
  conductance is given by 
  $\frac{e^2}{\pi\hbar}\big\langle\tr(tt^\dagger)\big\rangle$, taking into account two possible spin
  orientations; we  prefer to express the result in units of
  $\frac{e^2}{\pi\hbar}$),%
 \item the
{\it conductance variance} $\big\langle (\tr
\;tt^\dagger)^2\big\rangle-\big\langle
  \tr(tt^\dagger)\big\rangle^2$,
\item the mean {\it shot noise} power
  $\big\langle\tr(tt^\dagger-tt^\dagger tt^\dagger)\big\rangle$, in
  units of $\frac{2e^3|V|}{\pi\hbar}$,
\item for a cavity with three leads, {\it correlations between the
    currents flowing from lead 1 to lead 2 and from lead 1 to lead 3}
  depending on the corresponding transition matrices
  $t^{(1\to2)},t^{(1\to 3)}$ as
$\big\langle\tr(t^{(1\to2)}{t^{(1\to2)}}^\dagger
  t^{(1\to3)}{t^{(1\to3)}}^\dagger)\big\rangle$,
\item the {\it conductance covariance} at two different energies which
  characterizes the so-called Ericson fluctuations.
\end{itemize}
Here the angular brackets signify  an average over an energy interval
 sufficient to smooth out the fluctuations of the
respective physical property. We will see that after such an averaging each of
these quantities takes a universal form in agreement with
random-matrix theory, without any need for an ensemble average. 
To show this, 
we shall express the transition amplitudes as sums over
trajectories as in (\ref{transition}). The above observables then turn into averaged sums over pairs
or quadruplets of trajectories, which will be evaluated according
to simple and universal diagrammatic rules. We shall first derive and exploit
these rules for the orthogonal and unitary cases and then
generalize to the interpolating case (weak magnetic field).

Due to the unitarity of the time evolution,
we could equivalently express transport properties
through reflection amplitudes and trajectories starting and ending at the same lead.
For the average conductance we have checked explicitly that the same result is
obtained, meaning that our approach preserves unitarity.

\section{Mean conductance}

\label{sec:conductance}

We first consider the mean conductance and propose to show that
individual chaotic systems are faithful to the random-matrix
prediction \cite{Beenakker,Conductance}
\begin{equation}
\label{RMT_conductance}
\big\langle G(E)\big\rangle=\big\langle \tr(tt^\dagger)\big\rangle=
\begin{cases}
  \frac{N_1 N_2}{N} & \text{unitary case} \\
  \frac{N_1 N_2}{N+1} & \text{orthogonal case}\,.
\end{cases}
\end{equation}
In the semiclassical approximation (\ref{transition}), the average
conductance becomes a double sum over trajectories $\alpha,\beta$
connecting the same channels $a_1$ and $a_2$,
\begin{equation}
\label{doublesum}
\big\langle \tr(tt^\dagger)\big\rangle
=\left\langle \sum_{a_1,a_2}t_{a_1 a_2}t_{a_1 a_2}^*\right\rangle
=\frac{1}{T_H}\left\langle\sum_{a_1,a_2}\sum_{\alpha,\beta: a_1\to a_2}
A_\alpha A_\beta^*{\rm e}^{{{\rm i}}(S_\alpha-S_\beta)/\hbar}\right\rangle\,.
\end{equation}
Due to the phase factor ${\rm e}^{{{\rm i}}(S_\alpha-S_\beta)/\hbar}$, the
contributions of most trajectory pairs oscillate rapidly in the limit
$\hbar\to 0$, and vanish after averaging over the energy. Systematic
contributions can only arise from pairs with action differences
$\Delta S=S_\alpha-S_\beta$ of the order of $\hbar$.

\subsection{Diagonal contribution}

The simplest such pairs involve {\it identical trajectories}
$\alpha=\beta$, with a vanishing action difference \cite{Berry,
  Baranger}. These ``diagonal'' pairs contribute
\begin{equation}
\big\langle\tr(tt^\dagger)\big\rangle\big|_{\rm diag}
=\frac{1}{T_H}\left\langle\sum_{a_1,a_2}
\sum_{\alpha:a_1\to a_2}|A_\alpha|^2\right\rangle \,.
\end{equation}
The foregoing single-trajectory sum may be evaluated using the
following rule established by Richter and Sieber \cite{RS}: Summation
over trajectories connecting fixed channels is equivalent to
integration over the dwell time $T$ as
\begin{equation}\label{RSsum}
\sum_{\alpha:a_1\to a_2}|A_\alpha|^2=\int_0^\infty dT {\rm e}^{-\frac{N}{T_H}T}
=\frac{T_H}{N} \,.
\end{equation}
Here, the integrand ${\rm e}^{-\frac{N}{T_H}T}$ can be understood as
the {\it survival probability}, i.e., the probability for the
trajectory to remain inside the cavity up to the time $T$.  The factor
$\frac{N}{T_H}=\frac{2p(w_1+w_2)}{\Omega}$ is the classical escape
rate. Due to $\Omega\propto L^2$,
that rate is proportional to $\frac{w}{L}$ if $p$ is scaled according to
$p\propto L$; inversion yields the typical dwell
time $T_D=\frac{T_H}{N}\propto \frac{L}{w}$ mentioned in the introduction.

Finally summing over all $N_1$ possible choices for $a_1$ and over
the $N_2$ possibilities for $a_2$, one finds \cite{Baranger,RS}
\begin{equation}
\label{conductance_diagonal}
\big\langle\tr(tt^\dagger)\big\rangle\big|_{\rm diag}=\frac{N_1N_2}{N}\,.
\end{equation}
Eq. (\ref{conductance_diagonal}) reproduces the RMT result for the
unitary case, and gives the leading term in the
orthogonal case.

\subsection{Richter/Sieber pairs}

\label{sec:RS}
\begin{figure}
\begin{center}
\includegraphics[scale=1.0]{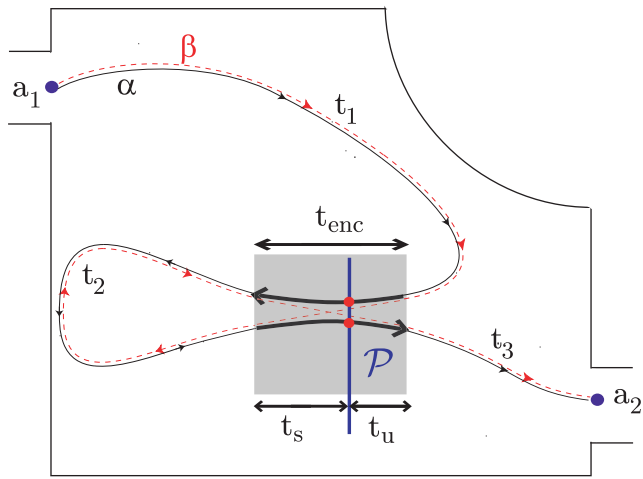}
\end{center}
\caption{Scheme of a Richter/Sieber pair. The trajectories
  $\alpha$ (full line) and $\beta$ (dashed line)
  connect the same channels $a_1$ and $a_2$, and differ only inside a
  2-encounter (in the box).  A Poincar{\'e} section ${\cal P}$
  intersects the encounter stretches at the points ${\bf x}_{\p 1}$
  and   ${\bf x}_{\p 2}$ in phase space whose configuration-space location is
  highlighted by two dots.
  ${\cal P}$ divides the encounter in two parts with durations
  $t_s\sim\frac{1}{\lambda}\ln\frac{c}{|s|}$ and
  $t_u\sim\frac{1}{\lambda}\ln\frac{c}{|u|}$.
  The relevant trajectories are in reality much longer than depicted here; in the absence
  of a potential they consist of a huge number of straight segments reflected at
  the boundary.
  }
\label{fig:RS}
\end{figure}

For the orthogonal case, Richter and Sieber attributed the
next-to-leading order to another family of trajectory pairs. In
the following, we shall describe these pairs in a language adapted
to an extension to higher orders in $\frac{1}{N}$. In each
Richter/Sieber pair (see Fig.~\ref{fig:RS}), the trajectory
$\alpha$ contains a
  ``2-encounter'' wherein two stretches are almost mutually
time-reversed; in configuration space it looks like either a
small-angle self-crossing  or a narrow avoided crossing. We demand
that these two stretches come sufficiently close such that their
motion is mutually linearizable.  Along $\alpha$, the two
stretches are separated from each other and from the leads by
three ``links''
\footnote{
In our previous papers \cite{EssenFF,Tau3,JapanEssen} we used the term ``loop'' to refer to
the comparatively long
orbit pieces connecting encounter stretches to one another or to
the openings. We decided to replace this expression by ``link''
which is more appropriate since the beginning and the
end of such a piece may be far removed from each other (in the
case of the initial and final link and of links between different
encounters).
}.
 The partner trajectory $\beta$ is distinguished from
$\alpha$ only by differently connecting these links inside the
2-encounter. Along the links, however, $\beta$ is practically
indistinguishable from $\alpha$; in particular, the entrance and
exit angles of $\alpha$ and $\beta$ (defined by the in- and out-channels) coincide
\footnote{Following Richter and Sieber we find the semiclassical
estimate for a conductance component
$\langle|t_{a_1a_2}|^2\rangle$ between two given in- and
out-channels and demand therefore that all contributing
trajectories have the same in- and out-angles. An alternative
 \cite{Brouwer} is to replace summation over channels in
the formula for the transport property by integration; then the
channel numbers of the contributing trajectories found through an
additional saddle point approximation will not be integer.}.
 The initial and final links are traversed in the same sense of
motion by $\alpha$ and $\beta$, while for the middle link the
velocities are opposite.  Obviously, such Richter/Sieber pairs
$\alpha,\beta$ can exist only in time-reversal invariant systems.
The two trajectories in a Richter/Sieber pair indeed have nearly
the same action, with the action difference originating mostly
from the encounter region.

We stress that inside a Richter/Sieber pair, the encounter
stretches and the leads must be {\it separated by links of
positive durations} $t_1,t_2,t_3>0$.
For the ``inner'' loop with duration $t_2$ the reasons were worked out in
previous publications dealing with periodic orbits (\cite{Mueller,HigherDim}, and 
\cite{EssenFF,MuellerThesis} for more complicated encounters):
Essentially, $\beta$ is obtained from $\alpha$ by switching connections
between four points where the encounter stretches begin and end;
to have four such points the stretches must be separated by a non-vanishing link.
The fact that the duration of the initial and final links is non-negative (the
encounter does not ``stick out'' through any of the openings) is
trivial in the case of the Richter/Sieber pair: Since the
encounter stretches are 
almost antiparallel a trajectory with an encounter ``sticking out''
would enter and exits the cavity
through the same opening and thus be irrelevant for the conductance.

Encounters have an important effect on the {\it survival
probability} \cite{EssenCond}.  The trajectory $\alpha$ is exposed
to the ``danger'' of getting lost from the cavity only during the
three links and on the first stretch of the encounter.  If the
first stretch remains inside the cavity, the second stretch, being
close to the first one (up to time reversal) must remain inside as
well. If we denote the duration of one encounter stretch by
$t_{\rm enc}$, the total ``exposure time'' is thus given by
$T_{\rm
  exp}=t_1+t_2+t_3+t_{\rm enc}$; it is shorter than the dwell time $T$
which includes a second summand $t_{\rm enc}$ representing the second
encounter stretch. Consequently, the survival probability ${\rm
  e}^{-\frac{N}{T_H}T_{\rm exp}}$ exceeds the naive estimate ${\rm
  e}^{-\frac{N}{T_H}T}$. In brief, encounters hinder the loss of a
trajectory to the leads.

To describe the {\it phase-space geometry} of a 2-encounter, we
consider a Poincar{\'e} section ${\cal P}$ orthogonal to the first
encounter stretch in an arbitrary phase-space point ${\bf x}_{\p
1}$. This section must also intersect the second stretch in a
phase-space point ${\bf x}_{\p 2}$ almost time-reversed with
respect to ${\bf x}_{\p 1}$. 
In Fig. \ref{fig:RS} the configuration-space locations of ${\bf x}_{\p 1}$
and ${\bf x}_{\p 2}$ are highlighted by two dots. 
For a hyperbolic, quasi
two-dimensional\footnote{ Our treatment can easily
  be extended to $f>2$, see
   \cite{HigherDim} and  the Appendices of
  \cite{EssenFF,MuellerThesis}.}
system, the small phase space separation between the time-reversed
${\cal T}{\bf x}_{\p 2}$ of ${\bf x}_{\p 2}$ and ${\bf x}_{\p 1}$
can be decomposed as \cite{Gaspard,Spehner,Turek}
\begin{equation}
{\cal T}{\bf x}_{\p 2}-{\bf x}_{\p 1}=s{\bf e}^s({\bf x}_{\p
1})+u{\bf e}^u({\bf x}_{\p 1})\,,
\end{equation}
where ${\bf e}^s({\bf x}_{\p 1})$ and ${\bf e}^u({\bf x}_{\p 1})$
are the so-called stable and unstable directions at ${\bf x}_{\p
1}$. If ${\cal P}$ moves along the trajectory, following the time
evolution of ${\bf x}_{\p 1}$, the unstable component $u$ will
grow exponentially while the stable component $s$ shrinks
exponentially. For times large compared with the ballistic time
($L/v$ with $v$ the velocity) the rate of growth (or shrinking) is
given by the Lyapunov exponent $\lambda$ (not to be confused with
the wavelength also denoted by $\lambda$),
\begin{eqnarray}
\label{asymptotics}
u(t)&\sim& u(0){\rm e}^{\lambda t}\nonumber\\
s(t)&\sim& s(0){\rm e}^{-\lambda t}\,.
\end{eqnarray}

By our definition of a 2-encounter, the stable and unstable
components are confined to ranges $-c<s<c$, $-c<u<c$, with $c$ a
small phase-space separation.
The exact value of $c$ will be irrelevant, except that the
transverse size of the encounter in configuration space, $\sim
c/\sqrt{m\lambda}$ with $m$ the mass, must be small compared with
the opening diameters. It should also be small enough to allow
mutual linearization of motion along the encounter stretches.
 As a consequence, the time between
${\cal P}$ and the end of the encounter is
$t_u\sim\frac{1}{\lambda}\ln\frac{c}{|u|}$, i.e., the time the
unstable component needs to grow from $u$ to $\pm c$. Likewise the
time between the beginning of the encounter and ${\cal P}$ reads
$t_s\sim\frac{1}{\lambda}\ln\frac{c}{|s|}$. Both times sum up to
the encounter duration
\begin{equation}
\label{duration} t_{\rm
enc}=t_u+t_s\sim\frac{1}{\lambda}\ln\frac{c^2}{|su|}\,.
\end{equation}
A glance at Fig.~\ref{fig:RS} shows that the times $t_{\p  1},t_{\p 
2}$ of the piercing points ${\bf x}_{\p 1}$, ${\bf x}_{\p 2}$
(measured from the beginning of the trajectory) are now given by
\begin{eqnarray}
\label{piercing_times} t_{\p  1}=t_1+t_u,\quad
 t_{\p 2}=t_1+t_{\rm enc}+t_2+t_s\,.
\end{eqnarray}
Finally, the stable and unstable coordinates determine the action
difference as \cite{Spehner,Turek} (see also \cite{Mueller,EssenFF})
\begin{equation}
\label{action_difference}
\Delta S=su\,.
\end{equation}
The  encounters relevant for the transport phenomena have action
differences of order $\hbar$ and thus durations $t_{\rm
  enc}\sim\frac{1}{\lambda}\ln\frac{c^2}{|\Delta S|}$ of the order of
the Ehrenfest time $T_E=\frac{1}{\lambda}\ln\frac{c^2}{\hbar}$.

With this input, we can determine the  average number of
  2-encounters inside trajectories $\alpha$ of a given dwell time
$T$. In ergodic systems, the probability for a trajectory to
pierce through a fixed Poincar{\'e} section ${\cal P}$ in a time
interval $(t_{\p  2},t_{\p  2}+dt_{\p  2})$ with stable and unstable
separations from ${\bf x}_{\p 1}$ inside $(s,s+ds)\times(u,u+du)$
is uniform, and given by the Liouville measure
$\frac{1}{\Omega}dt_{\p  2} ds du$. To count all 2-encounters
inside $\alpha$, we have to integrate this density over $t_{\p  2}$
(to get all piercings ${\bf x}_{\p 2}$ through a given ${\cal P}$)
and over $t_{\p  1}$ (to get all possible ${\bf x}_{\p 1}$ and thus
all possible sections ${\cal P}$). When integrating over $t_{\p 
1}$, we weigh the contribution of each encounter with the
corresponding duration $t_{\rm
  enc}$, since the section ${\cal P}$ may be placed at any point
within the encounter; therefore we must subsequently divide by $t_{\rm
  enc}$. The integration over the piercing times $t_{\p  1}$, $t_{\p  2}$
may be replaced by integration over the link durations $t_1,t_2$,
which as we stressed, must be positive; in addition
$t_3=T-t_1-t_2-2t_{\rm
  enc}$ must also be positive. Altogether,
we  obtain the following density of stable and unstable
coordinates,
\begin{equation}
w(s,u)=\int_{t_1,t_2>0\atop t_1+t_2<T-2t_{\rm enc}} dt_1 dt_2\;
\frac{1}{\Omega\,t_{\rm enc}(s,u)}\,.
\end{equation}
This density is normalized such that integration over all
$s,u$ belonging to a given interval of $\Delta S=su$ yields the number
of 2-encounters of $\alpha$ giving rise to action differences within
that interval.

To find what Richter/Sieber pairs contribute to the conductance
(\ref{doublesum}), we replace the sum over $\beta$ by a sum over
2-encounters inside $\alpha$ or, equivalently, an integral over
$w(s,u)$. The additional approximation
\footnote{ For long trajectories, the derivative
$\left(\frac{\partial \vartheta_2}{\partial y_1}\right)_{\vartheta_1}$
in $A_\alpha$ \cite{Richter} is proportional to the so-called
stretching factor $\Lambda_\alpha$, i.e., the factor by which an
initial separation along the unstable direction grows until the
end of the trajectory $\alpha$. This factor can be written as a
product of the (time-reversal invariant) stretching factors of the
individual links and encounter stretches. Since $\beta$ contains
practically the same links and stretches, we have
$\Lambda_\beta\approx\Lambda_\alpha$. All other factors almost
coincide as well (see \cite{Mueller,Turek,MuellerThesis} for the
Maslov index), entailing $A_\beta\approx A_\alpha$.}
$A_\beta\approx A_\alpha$ yields
\begin{equation}
\big\langle\tr(tt^\dagger)\big\rangle\big|_{\rm RS}
=\frac{1}{T_H}\left\langle\sum_{a_1,a_2}\int ds du
\sum_{\alpha:a_1\to a_2}|A_\alpha|^2w(s,u){\rm e}^{{{\rm i}} su/\hbar}\right\rangle\,.
\end{equation}
Next, we employ the Richter/Sieber rule to do the sum over
$\alpha$ by integrating over the dwell time $T$, with the integrand
involving the (modified) survival probability ${\rm
  e}^{-\frac{N}{T_H}T_{\rm exp}}={\rm
  e}^{-\frac{N}{T_H}(t_1+t_2+t_3+t_{\rm enc})}$.  The integral over
$T$ may be transformed into an integral over the duration of the
final link.  Moreover, summation over all channels $a_1=1\ldots
N_1$, $a_2=1\ldots N_2$ yields a factor $N_1N_2$. We are thus led
to
\begin{equation}
\label{RS_integral}
\big\langle\tr(tt^\dagger)\big\rangle\big|_{\rm RS}=\frac{N_1N_2}{T_H}\left\langle\int_0^\infty dt_1 dt_2 dt_3
\int ds du\frac{1}{\Omega\,t_{\rm enc}(s,u)}{\rm e}^{-\frac{N}{T_H}(t_1+t_2+t_3+t_{\rm enc}(s,u))}{\rm e}^{{{\rm i}} su/\hbar}\right\rangle\,.
\end{equation}
This integral factors into three independent integrals over the
link durations,
\begin{equation}
\label{link_integral} \int_0^\infty dt_i{\rm
e}^{-\frac{N}{T_H}t_i}=\frac{T_H}{N}\,,
\end{equation}
and an integral over the stable and unstable separations within the
encounter,
\begin{equation}
I=\left\langle\int ds du\frac{1}{\Omega\,t_{\rm enc}(s,u)}
{\rm e}^{-\frac{N}{T_H}t_{\rm enc}(s,u)}
{\rm e}^{{{\rm i}} su/\hbar}\right\rangle\,.
\end{equation}
The encounter integral $I$ can be evaluated if we expand the
exponential as ${\rm e}^{-\frac{N}{T_H}t_{\rm
    enc}(s,u)}=1-\frac{N}{T_H}t_{\rm enc}(s,u)+ \ldots$. As shown in
\cite{EssenFF}, the constant term yields $\big\langle ds
du\frac{1}{\Omega\,t_{\rm enc}(s,u)}{\rm e}^{{{\rm i}} su/\hbar}\big\rangle
=\big\langle\frac{2\lambda\hbar}{\Omega}\sin\frac{c^2}{\hbar}\big\rangle$
which oscillates rapidly as $\hbar\to 0$ and therefore vanishes after
averaging.  In the semiclassical limit, the value of $I$ is solely
determined by the linear term for which the denominator $t_{\rm
  enc}(s,u)$ cancels out,
\begin{equation}
\label{encounter_integral}
I=-\frac{N}{\Omega T_H}\left\langle\int ds du
   {\rm e}^{{{\rm i}} su/\hbar}\right\rangle
=-\frac{N}{T_H^2}\,,
\end{equation}
where we use$\int ds du{\rm e}^{{{\rm i}} su/\hbar}\to 2\pi\hbar$ and
$T_H=\frac{\Omega}{2\pi\hbar}$; all further terms vanish compared
to the linear one like $\frac{Nt_{\rm enc}}{T_H}\sim
\frac{T_E}{T_D}$ and may thus be neglected, given our previous
definition of the semiclassical limit. Since all occurrences of
$T_H$ in Eqs. (\ref{RS_integral}), (\ref{link_integral}) and
(\ref{encounter_integral}) mutually cancel, we can formulate the
following ``diagrammatic rule'': {\it Each link yields a factor
$\frac{1}{N}$ and each encounter a factor $-N$.} The result still
has to be multiplied with the number of channel combinations
$N_1N_2$. Altogether, the contribution of Richter/Sieber pairs to
the average conductance is hence determined as
\begin{equation}
\big\langle\tr(tt^\dagger)\big\rangle\big|_{\rm RS}=-\frac{N_1N_2}{N^2} \,.
\end{equation}

Our present treatment differs from Richter's and Sieber's original
paper \cite{RS} in two points: First, one has to exclude
encounters which stick out of the opening.  (In \cite{RS},
encounters were described through self-crossings in configuration
space, and only that crossing, approximately in the center of the
encounter, had to be located inside the cavity.) Second, one must
take into account that encounters hinder the escape into the
leads. While these two corrections mutually cancel for
Richter/Sieber pairs they will presently turn out of crucial
importance for higher-order contributions to the mean conductance,
as well as for other observables like shot noise.

\subsection{Diagrammatic rules for all orders in $\frac{1}{N}$}

\label{sec:conductance_general}

\begin{figure}
\begin{center}
\includegraphics[scale=0.9]{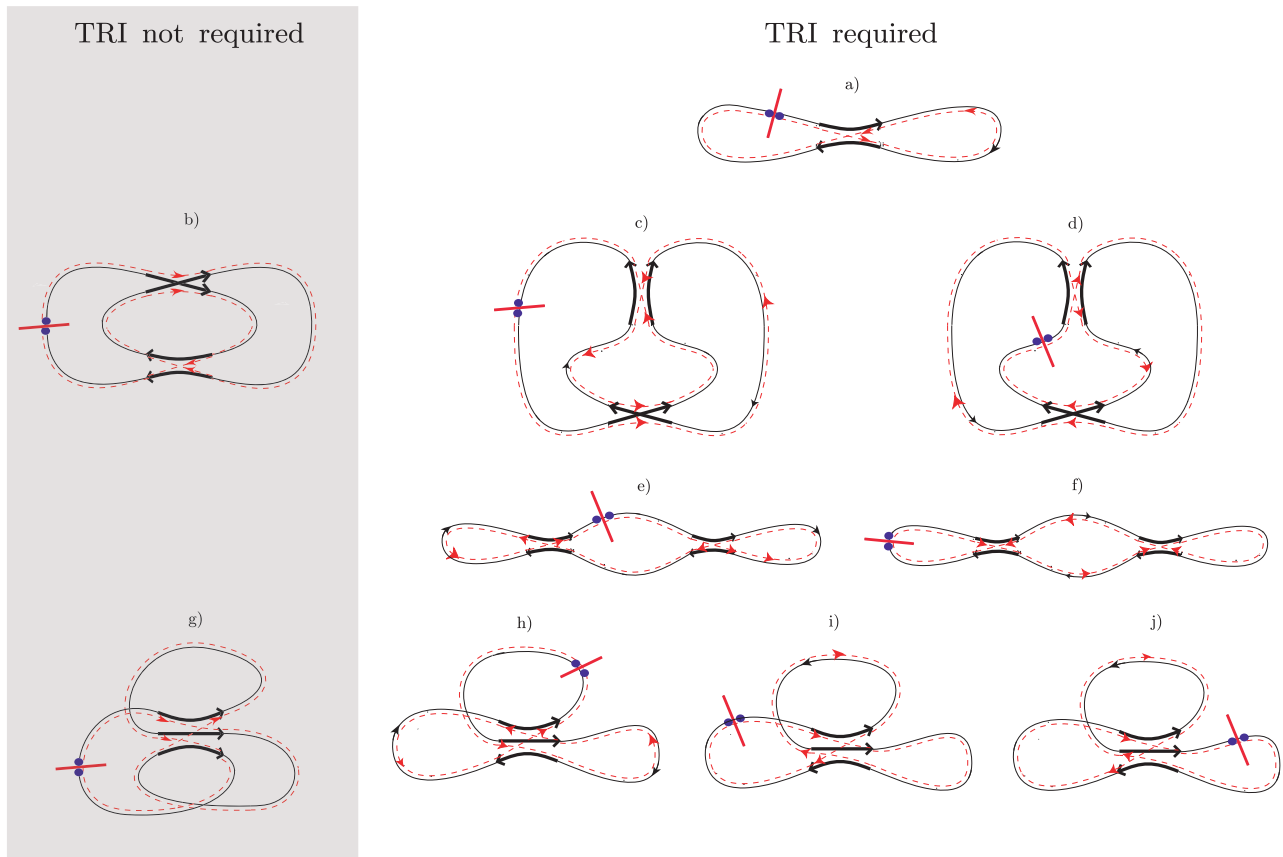}
\end{center}
\caption{ Families of trajectory pairs $(\alpha,\beta)$ differing in
  one 2-encounter (a), two 2-encounters (b-f) or in one 3-encounter
  (g-j). In contrast to Fig.~\ref{fig:RS} the cavity is not depicted,
  and the initial and final points of the trajectories are joined
  together (junction symbolized by two dots and intervening bar, $\cdot
  |\cdot$). One orbit pair may result from joining beginning and end of
  different trajectory pairs (like cd, ef, and hij).  Arrows indicate
  the directions of motion inside the encounters, and highlight those
  links which are traversed by $\alpha$ and $\beta$ with opposite sense of
  motion.  Note that all families apart from b) and g) involve almost
  mutually time-reversed encounter stretches and thus require
  time-reversal invariance (TRI). Together with the diagonal pairs,
  the ones shown here reproduce the mean conductance up to order
  $N_1N_2/N^3$.} \label{fig:conductance_next}
\end{figure}

\begin{figure}
\begin{center}
\includegraphics[scale=1.0]{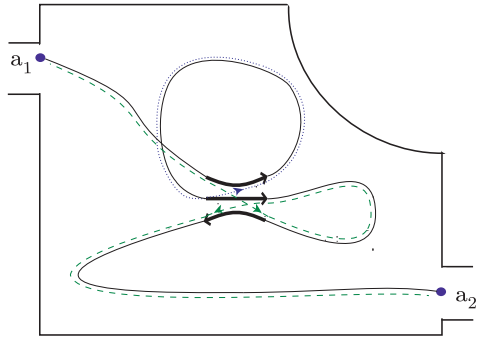}
\end{center}
\caption{ Reconnections inside a 3-encounter leading to one
  trajectory (dashed) and one periodic orbit (dotted), rather than a
  single connected partner trajectory.}
\label{fig:conductance_decompose}
\end{figure}

To proceed to all orders in $\frac{1}{N}$, we must consider pairs
of trajectories differing by their connections inside {\it
arbitrarily many encounters}. Each of these encounters may involve
{\it arbitrarily many stretches}.
 We shall speak of an $l$-encounter whenever
$l$ stretches of a trajectory come close in phase space. In
time-reversal invariant systems we must also allow for encounters
whose stretches are almost mutually time-reversed. As a
consequence, the resulting conductance will depend on the symmetry
class: We shall see all higher-order contributions to mutually
cancel in the unitary case but to yield the $1/N$ expansion of the
RMT result (\ref{RMT_conductance}) in the orthogonal case.

A list of examples is displayed in Fig.
\ref{fig:conductance_next}, where for later convenience we did not
draw the cavity and formally joined the initial and final points
of the trajectories together.  These examples illustrate the
simplest among infinitely many topologically different {\it
families of
  trajectory pairs}.

The individual families are characterized (i) by the {\it
  numbers $v_l$ of $l$-encounters} in which the two partners differ.
These numbers can be assembled into a
``vector''$\vec{v}=(v_2,v_3,v_4,\ldots)$, and determine the
overall number of encounters $V(\vec{v})=\sum_{l\geq 2}v_l$ and
the total number of encounter stretches $L(\vec{v})=\sum_{l\geq
2}lv_l$. The number of links exceeds the number of stretches by
one and reads $L(\vec{v})+1$.  Further characteristics of our
families of trajectory pairs are (ii) the {\it order in which the
encounters are traversed by the trajectory $\alpha$}, (iii) the
{\it mutual orientation of the encounter stretches} (i.e., $\psr$
vs. $\asr$, or $\pc$ vs.  $\ac$), and (iv) the {\it reconnections}
leading to the partner trajectory $\beta$.  We stress that $\beta$
must be a single connected trajectory; reconnections leading to,
e.g., one trajectory and one periodic orbit as in Fig.
\ref{fig:conductance_decompose} must be excluded.

All families of trajectory pairs contribute to the conductance
according to the {\it same rules as do Richter/Sieber pairs}: Each
link yields a factor $\frac{1}{N}$, each encounter gives a factor
$-N$, and we have to multiply with the number of channel
combinations $N_1N_2$. To prove this assertion, we place a
Poincar{\'e} section ${\cal
  P}_\sigma$ (with $\sigma=1\ldots V$) across each of the $V$
encounters. Similar as for Richter/Sieber pairs, we
characterize each $l$-encounter by $l-1$ stable coordinates $s_{\sigma
  j}$ (with $\sigma=1\ldots V$ and $j=1\ldots l-1$), and by $l-1$
unstable coordinates $u_{\sigma j}$ \cite{EssenFF}, measuring the
phase-space separations between the points where the $l$ encounter
stretches pierce through ${\cal P}_\sigma$.

All $V$ encounters are thus characterized by $\sum_{l\geq
2}(l-1)v_l=L-V$ stable coordinates, and the same number of unstable
coordinates. As shown in \cite{EssenFF}, these coordinates determine
the action difference as $\Delta S=\sum_{\sigma,j}s_{\sigma
j}u_{\sigma j}$.  Each encounter lasts as long as the absolute
values of all coordinates remain below the bound $c$.  Consequently,
the duration of an encounter is determined by the largest stable and
unstable coordinates, the first to reach $c$. In analogy to
(\ref{duration}), the $\sigma$-th encounter thus has the duration
\begin{equation}
t_{\rm enc}^\sigma(s,u)\sim\frac{1}{\lambda}
\ln\frac{c^2}{\max_j|s_{\sigma j}|\times
\max_{j'}|u_{\sigma j'}|}\,.
\end{equation}

Again, the trajectory $\alpha$ may get lost from the cavity only
during the links and on the first stretch of each encounter. If it
survives on that first stretch, it cannot escape on the remaining
stretches of the same encounter, since these are close to the
first. The exposure time $T_{\rm exp}$ is thus obtained as the sum
of all link and encounter durations $T_{\rm
exp}=\sum_{i=1}^{L+1}t_i+\sum_{\sigma=1}^V t_{\rm enc}^\sigma$.
The {\it survival probability} ${\rm e}^{-\frac{N}{T_H}T_{\rm
exp}}$ results,  again larger than the naive estimate ${\rm
e}^{-\frac{N}{T_H}T}$.

We proceed to investigating the {\it statistics} of encounters {\it
  for a given family} of trajectory pairs. Generalizing the treatment
of Subsection \ref{sec:RS} we first keep all $V$ Poincar{\'e}
sections ${\cal P}_\sigma$ fixed: each ${\cal P}_\sigma$ is placed
orthogonal to $\alpha$ at a phase-space point traversed at a fixed
time. The further $L-V$ piercings through these sections may be
considered statistically independent; the probability density for
their occurrence at given times and with given stable and unstable
coordinates thus reads $\frac{1}{\Omega^{L-V}}$. To account for
all encounters, we must integrate $\frac{1}{\Omega^{L-V}}$ over
all times for the $L-V$ later piercings. Moreover, to include all
possible ${\cal P}_\sigma$, we must integrate over the times of
the $V$ phase-space points chosen as the origin of a section
${\cal P}_\sigma$.  Since all sets of piercings within the
duration $t_{\rm enc}^\sigma$ belong to the same encounter, the
latter integral weighs each encounter with a factor $t_{\rm
enc}^\sigma$. Exactly as for Richter/Sieber pairs, this factor
must subsequently be divided out.  To simplify our calculation, we
replace the integral over altogether $L$ times of piercings by an
integral over the durations $t_1,t_2,\ldots,t_L$ of all links
except the final one. This replacement is permissible because all
piercing times may be written as functions of
$t_1,t_2,\ldots,t_L$, with the Jacobian of the transformation
equal to unity. Here, the duration $t_{L+1}$ of the final link
does not show up, since that link does not precede any encounter
stretch or piercing point.  The integral goes over positive $t_i$,
by the same reasoning as for 2-encounters
\footnote{
Note that the initial and final links must have positive duration
also in case they lead to stretches of a parallel encounter. Otherwise all
parallel stretches of that encounter would enter or leave the cavity
through the same lead. This 
is impossible, since the trajectories cannot enter or leave the cavity
several times.}; 
moreover, the
cumulative duration of the first $L$ links and all encounter
stretches must be smaller than the dwell time $T$, to allow for a
non-vanishing final link. We thus obtain the following density of
stable and unstable coordinates
\begin{equation}
\label{w_conductance_general}
w(s,u)=\int_{t_i>0\atop \sum_{i=1}^L t_i+\sum_{\sigma=1}^V
t_{\rm enc}^\sigma<T} dt_1\ldots dt_L
\;\frac{1}{\Omega^{L-V}\prod_{\sigma=1}^Vt_{\rm enc}^\sigma(s,u)}\,.
\end{equation}
The weight $w(s,u)$ is normalized similarly as in Subsection
\ref{sec:RS}: Integration over $s_{\sigma j},u_{\sigma j}$
corresponding to a given interval of $\Delta
S=\sum_{\sigma,j}s_{\sigma j}u_{\sigma j}$ leads to the number of
partner trajectories $\beta$ of a given $\alpha$, with action
difference inside that interval, and with the pair $(\alpha,\beta)$
belonging to the family considered.

We can now evaluate the {\it contribution of an arbitrary family
of trajectory pairs} to the average conductance (\ref{doublesum}).
We again approximate $A_\beta\approx A_\alpha$, and write the sum
over $\beta$ as an integral over $w(s,u)$,
\begin{equation}
\big\langle\tr(tt^\dagger)\big\rangle\big|_{\rm fam}
=\frac{1}{T_H}\left\langle\sum_{a_1,a_2}\int d^{L-V}s\;
d^{L-V}u\sum_{\alpha:a_1\to a_2}|A_\alpha|^2 w(s,u)
{\rm e}^{{\rm i}\sum_{\sigma,j}s_{\sigma j}u_{\sigma j}/\hbar}\right\rangle \,.
\end{equation}
As before, the sum over $a_1,a_2$ leads to a multiplication with
the number of channel combinations $N_1N_2$.  The sum over
$\alpha$ can be done using the (modified) Richter/Sieber sum rule,
and leads to integration over the duration $t_{L+1}$ of the final
link, with an integrand involving the survival probability ${\rm
  e}^{-\frac{N}{T_H}T_{\rm exp}}={\rm
  e}^{-\frac{N}{T_H}(\sum_{i=1}^{L+1}t_i+\sum_{\sigma=1}^V t_{\rm
    enc}^\sigma)}$.  Together with the integrals over the remaining
$L$ link durations in $w(s,u)$, Eq. (\ref{w_conductance_general}),
all links are now treated equally, 
and we obtain

\begin{equation}
\label{conductance_factored}
\big\langle\tr(tt^\dagger)\big\rangle|_{\rm fam}
=\frac{N_1 N_2}{T_H}\prod_{i=1}^{L+1}
\left(\int dt_i {\rm e}^{-\frac{N}{T_H}t_i}\right)
\prod_{\sigma=1}^V\left(\int ds_{\sigma1}\ldots ds_{\sigma,l-1} du_{\sigma1}
\ldots du_{\sigma,l-1}
\frac{{\rm e}^{{\rm i}\sum_js_{\sigma j}u_{\sigma j}}
{\rm e}^{-\frac{N}{T_H}t_{\rm enc}^\sigma(s,u)}}
{\Omega^{l-1}t_{\rm enc}^\sigma(s,u)}\,
\right)\,.
\end{equation}
Just like in (\ref{RS_integral}) the integral factorizes into
several integrals, one for each link and each encounter: Each link
gives $\frac{T_H}{N}$, and the encounter integral is determined by
the linear term in the series expansion of the exponential ${\rm
e}^{-\frac{N}{T_H}t_{\rm enc}^\sigma(s,u)}$. The encounter
integral is slightly changed because the piercing probability
$\frac{1}{\Omega}$, and the simple integral $\int ds du{\rm
e}^{{{\rm i}} su/\hbar}\to 2\pi\hbar$ now both appear in the
$(l-1)$-fold power.  Each encounter thus yields
$-\frac{N}{T_H}\left(\frac{2\pi\hbar}{\Omega}\right)^{l-1}=-\frac{N}{T_H^l}$.
Again, all powers of $T_H$ mutually cancel. We thus find the same
diagrammatic rules as above with a factor $\frac{1}{N}$ from each
of the $L+1$ links, a factor $-N$ from each of the $V$ encounters,
and a factor $N_1N_2$ representing the possible channel
combinations. 
(We note that these rules have a nice analogy to our previous work on
spectral statistics, see Appendix \ref{sec:spectral}).
The contribution of each family is therefore given
by
\begin{equation}
\label{conductance_family}
\big\langle\tr(tt^\dagger)\big\rangle\big|_{\rm fam}
=(-1)^{V(\vec{v})}\frac{N_1 N_2}{N^{L(\vec{v})-V(\vec{v})+1}}\,.
\end{equation}

To obtain the {\it overall conductance}, we must sum over all
families. If we let ${\cal N}(\vec{v})$ denote the number of families
associated to $\vec{v}$, Eqs. (\ref{conductance_diagonal}) and
(\ref{conductance_family}) imply
\begin{equation}
\label{conductance_sum}
\big\langle\tr(tt^\dagger)\big\rangle=\frac{N_1 N_2}{N}
\left(1+\sum_{\vec{v}}(-1)^{V(\vec{v})}
\frac{1}{N^{L(\vec{v})-V(\vec{v})}}{\cal N}(\vec{v})\right)
=\frac{N_1 N_2}{N}\left(1+\sum_{m=1}^\infty \frac{c_m}{N^m}\right)\,.
\end{equation}
Here, each coefficient $c_m$ is determined by the families with given
$m=L-V$,
\begin{equation}
\label{cm}
c_m=\sum_{\vec{v}}^{L(\vec{v})-V(\vec{v})=m}(-1)^{V(\vec{v})}{\cal N}(\vec{v})\,.
\end{equation}
Our task has thus been reduced to counting families of trajectory
pairs and evaluating $c_m$.

For the {\it lowest orders $m$}, the counting is easily done. We
have already seen that for time-reversal invariant systems the
next-to-leading contribution to conductance originates from the
Richter/Sieber family of trajectory pairs differing in one
2-encounter, see Fig.~\ref{fig:RS} or \ref{fig:conductance_next}a.
This family has $L=2$, $V=1$ thus $m=1$. Hence, it gives
rise to a coefficient $c_1=-1$ in the orthogonal case. In the unitary
case the absence of such trajectory pairs implies $c_1=0$.

The following coefficient $c_2$ is determined by pairs of trajectories
which either differ in two 2-encounters (i.e., $L=4$, $V=2$) and thus
contribute with a positive sign, or differ in one 3-encounter (i.e.,
$L=3$, $V=1$) and contribute with a negative sign. The relevant
families are sketched in Fig.~\ref{fig:conductance_next}. In the
unitary case, there is only one family of the first type
(Fig.~\ref{fig:conductance_next}b), and one family of the second type
(Fig.~\ref{fig:conductance_next}g). Both contributions mutually
cancel, i.e., $c_2=0$. In the orthogonal case, we must allow for
encounters with mutually time-reversed stretches. We then find five
families contributing with a positive sign
(Fig.~\ref{fig:conductance_next}b-f) and four families contributing
with a negative sign (Fig.~\ref{fig:conductance_next}g-j). All
contributions sum up to $c_2=1$.

The higher coefficients $c_m$ require more involved combinatorial
methods to which we now turn.

\subsection{Combinatorics}\label{Combi}

\label{sec:conductance_combinatorics}

In \cite{EssenFF}, we found a systematic way for summing
contributions of families of trajectory pairs that differ in
arbitrarily many encounters. In particular, we obtained a
recursion for the numbers ${\cal N}(\vec{v})$. Since the treatment
of \cite{EssenFF} was geared towards spectral statistics of {\it
closed} systems, it was formulated for pairs of {\it periodic
orbits} rather than pairs of open trajectories.  It can, however,
be easily adapted to open trajectories.  We just have to turn
trajectory pairs into orbit pairs by joining the initial and final
points as in Fig.~\ref{fig:conductance_next}, or cut orbits in
order to form trajectories.

The topology of orbit pairs $({\cal A},{\cal B})$ was described by
{\it ``structures''}. To define these structures, we numbered the
encounter stretches of ${\cal A}$ in their order of traversal,
starting from an arbitrary reference stretch; the links of ${\cal
A}$ were numbered as well, with the first link preceding the first
encounter stretch. Then, each structure is characterized by (i) a
vector $\vec{v}$ as above,
 (ii) a way of distributing the numbered
stretches among the encounters, (iii) fixing the mutual orientation of stretches
inside each encounter, and (iv) shifting connections to form a
partner orbit ${\cal B}$.

With this definition, each family of trajectory pairs corresponds to
one structure of orbit pairs. We only have to glue together the
initial and final points of the trajectories as in
Fig.~\ref{fig:conductance_next}, and keep the first stretch of
$\alpha$ (the one following the initial point) as the first stretch of
${\cal A}$. Thus, each of the pictures in
Fig.~\ref{fig:conductance_next} represents one structure of orbit
pairs, and the numbers of orbit pair structures and of trajectory pair
families both equal ${\cal N}(\vec{v})$.

To illustrate this relation, we consider the two families of
trajectory pairs depicted in Fig.~\ref{fig:conductance_next}e and
f. If we join the initial and final points for any of these families,
we obtain orbit pairs of one and the same topology.  But still the
resulting structures are different, because in
Fig.~\ref{fig:conductance_next}e the first encounter stretch (the one
following the initial point of the trajectory) precedes a stretch of
the same encounter, i.e., the two antiparallel encounters respectively
involve the stretches $(1,2)$ and $(3,4)$.  A different choice of the
initial stretch as in Fig.~\ref{fig:conductance_next}f means that the
two encounters comprise the stretches $(1,4)$ and $(2,3)$. Indeed,
structures of orbit pairs and families of trajectory pairs are
one-to-one.

For later convenience, we refer to the structure involving one
antiparallel encounter (Fig.~\ref{fig:conductance_next}a) as the {\it
  Sieber/Richter structure} (it was proposed by these authors in
\cite{SR}, see also \cite{Fieldtheory}), to the structure involving
two {\it p}arallel 2-encounters (Fig.~\ref{fig:conductance_next}b) as
{\it ppi}, to the two structures involving a {\it p}arallel and an
{\it a}ntiparallel 2-encounter (Fig.~\ref{fig:conductance_next}c and
d) as {\it api}, and to the structures involving two {\it
  a}ntiparallel encounters (Fig.~\ref{fig:conductance_next}e and f) as
{\it aas} \cite{Tau3}.  The structure involving one {\it p}arallel
3-encounter (Fig.~\ref{fig:conductance_next}g) will be called {\it
  pc}, and the three structures of the type $\ac$
(Fig.~\ref{fig:conductance_next}h-j) {\it ac}.

To formulate our recursion for ${\cal N}(\vec{v})$, we now denote by
${\cal N}(\vec{v},l)$ the number of structures of orbit pairs (or
families of trajectory pairs) related to $\vec{v}$, {\it for which the
  first stretch belongs to an $l$-encounter}. We
had established the identity
\begin{equation}
\label{Nvl}
{\cal N}(\vec{v},l)=\frac{lv_l}{L(\vec{v})}{\cal N}(\vec{v})
\end{equation}
which has the following intuitive interpretation (not to be confused
with the proof in \cite{EssenFF}): The probability that the first
stretch forms part of an $l$-encounter is given by the overall number
$lv_l$ of stretches belonging to $l$-encounters, divided by the
overall number $L(\vec{v})$ of stretches in all encounters; to obtain
${\cal N}(\vec{v},l)$, we have to multiply ${\cal N}(\vec{v})$ with
that probability. Incidentally, the definition of ${\cal
  N}(\vec{v},l)$ implies $\sum_{l\geq 2}{\cal N}(\vec{v},l)={\cal
  N}(\vec{v})$.

The relevant recursion relation reads
\cite{EssenFF}\footnote{ Eq. (\ref{recursion2}) is a special case of
  Eqs. (42) and (54) in \cite{EssenFF}, with $l=2$. To understand the
  equivalence, note that in \cite{EssenFF} we  allowed for
  ``vectors'' $\vec{v}'$ including a non-vanishing component $v_1'$,
  which may formally be interpreted as a number of ``1-encounters''.
  We moreover showed that ${\cal N}(\vec{v}',1)={\cal
    N}(\vec{v}'^{[1\to]})$ (see the paragraph preceding Eq. (58) of
  \cite{EssenFF}). Applying this relation to $\vec{v}'=\vec{v}^{[2\to
    1]}$, one sees that the ${\cal N}(\vec{v}^{[2\to 1]},1)$ appearing
  in Eq. (54) of \cite{EssenFF} coincides with ${\cal
    N}(\vec{v}^{[2\to]})$.  }
\begin{equation}
\label{recursion2}
{\cal N}(\vec{v},2)-\sum_{k\geq 2}{\cal N}(\vec{v}^{[k,2\to k+1]},k+1)
=\left(\frac{2}{\beta}-1\right){\cal N}(\vec{v}^{[2\to]})\,,
\end{equation}
with $\beta=2$ and $\beta=1$ respectively referring to the unitary and
orthogonal case.  The symbol $\vec{v}^{[k,2\to k+1]}$ denotes the
vector obtained from $\vec{v}$ if we reduce $v_k$ and $v_2$ by one,
and increase $v_{k+1}$ by one; likewise $\vec{v}^{[2\to]}$ is obtained
from $\vec{v}$ if we reduce $v_2$ by one. In general, the list on the
left-hand side of the arrow contains the sizes of ``removed''
encounters, whereas the right-hand side contains the sizes of
``added'' encounters.

To turn (\ref{recursion2}) into a recursion for the coefficients
$c_m$, we multiply with $(-1)^{V(\vec{v})}$ and sum over all $\vec{v}$
with fixed $m=M(\vec{v})\equiv L(\vec{v})-V(\vec{v})$,
\begin{equation}
\label{recursion2_summed}
\sum_{\vec{v}}^{M(\vec{v})=m}(-1)^{V(\vec{v})}{\cal N}(\vec{v},2)
-\sum_{k\geq 2}\sum_{\vec{v}}^{M(\vec{v})=m}(-1)^{V(\vec{v})}
{\cal N}(\vec{v}^{[k,2\to k+1]},k+1)
=\left(\frac{2}{\beta}-1\right)\sum_{\vec{v}}^{M(\vec{v})=m}
(-1)^{V(\vec{v})}{\cal N}(\vec{v}^{[2\to]})\,.
\end{equation}
In each of the foregoing sums we can replace the summation variable
$\vec{v}$ by $\vec{v}'\equiv\vec{v}^{[k,2\to k+1]}$ or
$\vec{v}''\equiv\vec{v}^{[2\to]}$. Given that trajectory pairs
associated with $\vec{v}'$ have one encounter and one encounter
stretch less than those of $\vec{v}$, we then have to sum over
$\vec{v}'$ with
$M(\vec{v}')=L(\vec{v}')-V(\vec{v}')=(L(\vec{v})-1)-(V(\vec{v})-1)=m$.
By definition, we should restrict ourselves to $\vec{v}'$ with
$v'_{k+1}>0$; however, that restriction may be dropped since
$\vec{v}'$ with $v'_{k+1}=0$ have ${\cal N}(\vec{v}',k+1)=0$. In
contrast, trajectory pairs associated to $\vec{v}''$ have one
encounter and two stretches less than those associated to $\vec{v}$.
The pertinent sum runs over $\vec{v}''$ with
$M(\vec{v}'')=L(\vec{v}'')-V(\vec{v}'')=(L(\vec{v})-2)-(V(\vec{v})-1)=m-1$.
Using $(-1)^{V(\vec{v})}=-(-1)^{V(\vec{v}')}=-(-1)^{V(\vec{v}'')}$ we
can rewrite (\ref{recursion2_summed}) as
\begin{equation}
\label{recursion2_renamed}
\sum_{\vec{v}}^{M(\vec{v})=m}(-1)^{V(\vec{v})}{\cal N}(\vec{v},2)
+\sum_{k\geq 2}\sum_{\vec{v}'}^{M(\vec{v}')=m}(-1)^{V(\vec{v}')}
{\cal N}(\vec{v}',k+1)
=-\left(\frac{2}{\beta}-1\right)\sum_{\vec{v}''}^{M(\vec{v}'')=m-1}
(-1)^{V(\vec{v}'')}{\cal N}(\vec{v}'')\,.
\end{equation}
The left-hand side now boils down to
$\sum_{\vec{v}}^{M(\vec{v})=m}(-1)^{V(\vec{v})}\sum_{k\geq 1}{\cal
  N}(\vec{v},k+1) =\sum_{\vec{v}}^{M(\vec{v})=m}(-1)^{V(\vec{v})}{\cal
  N}(\vec{v}) =c_m$ while the right-hand side reads
$-\left(\frac{2}{\beta}-1\right)c_{m-1}$. We thus end up with a
recursion for $c_m$, $m\geq 2$,
\begin{equation}
\label{recursion_c}
c_m=-\left(\frac{2}{\beta}-1\right)c_{m-1}=\begin{cases}
0 & \mbox{unitary case}\\
-c_{m-1} & \mbox{orthogonal case}\,.
\end{cases}
\end{equation}

For the unitary case we conclude that all off-diagonal contributions
to the average conductance mutually cancel; the remaining diagonal
term, $\frac{N_1N_2}{N}$ reproduces the random-matrix result. For the
orthogonal case an initial condition is provided by the coefficient
$c_1=-1$, originating from Richter/Sieber pairs; hence $c_m=(-1)^m$.
The anticipated mean conductance (\ref{RMT_conductance}) is recovered
through (\ref{conductance_sum}) as the geometric series
\begin{equation}
\big\langle\tr(tt^\dagger)\big\rangle
=\frac{N_1N_2}{N}\left(1+\sum_{m=1}^\infty\frac{(-1)^m}{N^m}\right)
=\frac{N_1N_2}{N+1}\,.
\end{equation}
We have thus shown for both symmetry classes that the energy-averaged
conductance of individual chaotic cavities takes the universal form
predicted by random-matrix theory as an ensemble average.

\section{Conductance variance}

Experiments with chaotic cavities also reveal universal
conductance {\it fluctuations}. In particular, the conductance variance
agrees with the random-matrix prediction \cite{Beenakker}
\begin{equation}
\label{variance_RMT}
\big\langle G(E)^2\big\rangle-\big\langle G(E)\big\rangle^2
=\big\langle (\tr(tt^\dagger))^2\big\rangle
-\big\langle \tr(tt^\dagger)\big\rangle^2
=\begin{cases}
\frac{N_1^2N_2^2}{N^2(N^2-1)} & \mbox{unitary case}\\
\frac{2N_{1}N_{2}(N_{1}+1)(N_{2}+1)}{N(N+1)^{2}(N+3)}
& \mbox{orthogonal case\,.}
\end{cases}
\end{equation}
Once more, the semiclassical limit offers itself for an explanation of
such universality.  With the van Vleck approximation for the
transition amplitudes (\ref{transition}), the mean squared conductance
turns into a sum over quadruplets of trajectories,
\begin{eqnarray}
\label{quadsum}
\big\langle(\tr(tt^\dagger))^2\big\rangle
=\left\langle\sum_{a_1,c_1\atop a_2,c_2}t_{a_1a_2}t_{a_1a_2}^*
t_{c_1c_2}t_{c_1c_2}^*\right\rangle
=\frac{1}{T_H^2}\left\langle\sum_{a_1,c_1\atop a_2,c_2}
\sum_{\alpha,\beta:a_1\to a_2\atop\gamma,\delta: c_1\to c_2}
A_\alpha A_\beta^*A_\gamma A_\delta^*
{\rm e}^{{\rm i}(S_\alpha-S_\beta+S_\gamma-S_\delta)/\hbar}
\right\rangle\,.
\end{eqnarray}
Here $a_1,c_1=1\ldots N_1$ and $a_2,c_2=1\ldots N_2$ are channel
indices.  The trajectories $\alpha$ and $\beta$ lead from the same
ingoing channel $a_1$ to the same outgoing channel $a_2$, whereas
$\gamma$ and $\delta$ connect the ingoing channel $c_1$ to the
outgoing channel $c_2$. We can expect systematic contributions to the
quadruple sum over trajectories only from quadruplets with action
differences $\Delta S\equiv S_\alpha-S_\beta+S_\gamma-S_\delta$ of the
order of $\hbar$.

\subsection{Diagonal contributions}

The leading contribution to (\ref{quadsum}) originates from
``diagonal'' quadruplets with pairwise coinciding trajectories
either as $(\alpha=\beta$, $\gamma=\delta)$, or as
$(\alpha=\delta$, $\beta=\gamma$); both scenarios imply vanishing
action differences. The first scenario $\alpha=\beta$,
$\gamma=\delta$ obviously leads to $\beta$ connecting the same
channels as $\alpha$, and $\delta$ connecting the same channels as
$\gamma$, as required in (\ref{quadsum}); this holds regardless of
the channel indices $a_1,c_1,a_2,c_2$. The second scenario
($\alpha=\delta$, $\beta=\gamma$) brings about admissible
quadruplets only if all trajectories connect the same channels,
i.e., both the ingoing channels $a_1=c_1$ and the outgoing
channels $a_2=c_2$ coincide.  The contribution of these diagonal
quadruplets to (\ref{quadsum}) may thus be written as the
following double sum over trajectories $\alpha$ and $\gamma$
\begin{equation}
\label{quaddiagsum}
\big\langle(\tr(tt^\dagger))^2\big\rangle\big|_{\rm diag}=
\frac{1}{T_H^2}\left\langle\sum_{a_1,c_1\atop a_2,c_2}\sum_{\alpha
=\beta:a_1\to a_2\atop\gamma=\delta:c_1\to c_2}|A_\alpha|^2
|A_\gamma|^2
+\sum_{a_1=c_1\atop a_2=c_2}\sum_{\alpha=\delta:a_1\to a_2\atop \gamma
=\beta:a_1\to a_2}|A_\alpha|^2|A_\gamma|^2\right\rangle\,.
\end{equation}
The sum over channels just yields the number of possible channel
combinations as a factor, namely $N_1^2N_2^2$ for the first scenario
and $N_1N_2$ for the second one. Doing the sums over $\alpha$ and
$\gamma$ with the Richter/Sieber rule (\ref{RSsum}) we get
\begin{equation}
\label{quaddiag}
\big\langle(\tr(tt^\dagger))^2\big\rangle\big|_{\rm diag}=
\frac{N_1^2N_2^2+N_1N_2}{N^2}\,.
\end{equation}
The larger one of the two summands,
$\frac{N_1^2N_2^2}{N^2}$, is  cancelled by the
squared diagonal contribution to the mean conductance.

In recent works on Ehrenfest-time corrections
\cite{Brouwer,Whitney} (which are vanishingly small in our limit
$T_E\ll T_D$) the diagonal approximation was extended to include
trajectories which slightly differ close to the openings. The
relation of the methods used in these papers to our present
approach is not fully settled yet; further investigation about
this relation is desirable.

\subsection{Trajectory quadruplets differing in encounters}

\label{sec:quadruplets}

\begin{figure}
\begin{center}
\includegraphics[scale=0.9]{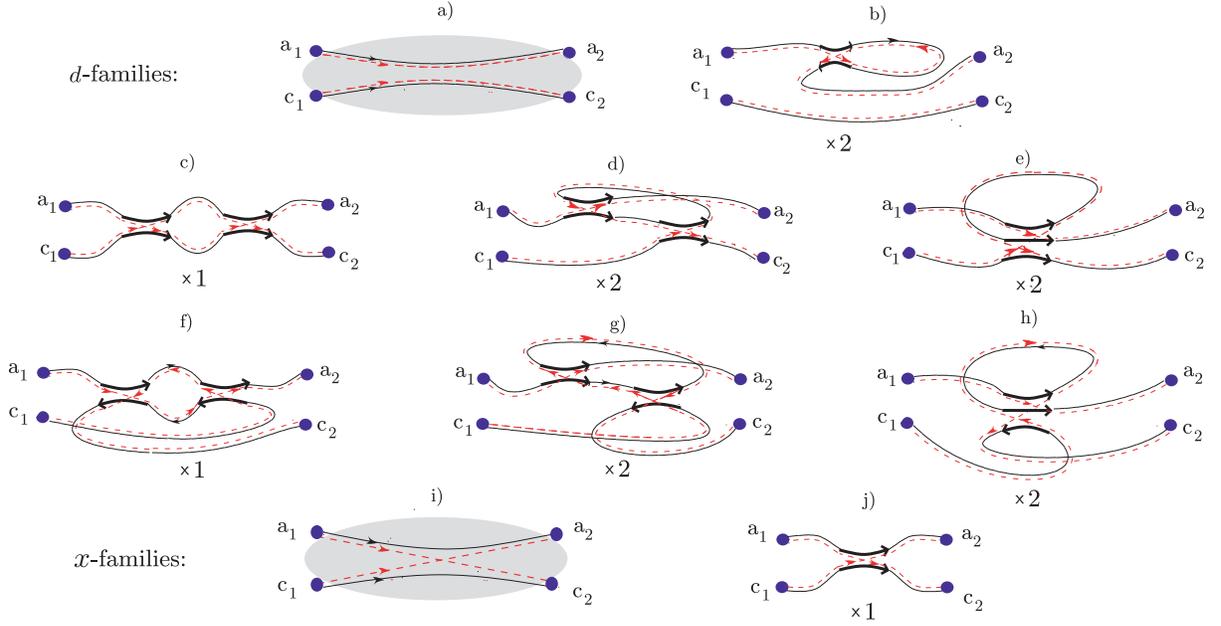}
\end{center}

  \caption{(a) Schematic graph of {\it d-quadruplets}, with the hatched
  area a ``black box'' containing any number of encounters; one of the
  dashed partner trajectories shares initial and final points with
  $\alpha$; the second partner trajectory similarly related to
  $\gamma$.\\
  (b)-(h)  d-quadruplets responsible for the leading-order
  contribution to the conductance variance. The diagrams (b), (f)-(h) containing encounters
   with antiparallel stretches exist only
  in the orthogonal case.
  A diagram may have a ``twin'' obtained by reflection  in a horizontal
  line; the number of  symmetric versions of each
  diagram is indicated by a multiplier underneath.\\
  (i) Schematic graph of {\it $x$-quadruplets},
  encounters suppressed: one of the partners shares initial link with
  $\gamma$ and final link with $\alpha$, the second one connects
  initial link of $\alpha$ with final link of $\gamma$. (j) An
  $x$-quadruplet involving one 2-encounter.  }
\label{fig:quadruplets}
\end{figure}

Off-diagonal contributions arise from quadruplets of trajectories
differing in encounters; see Fig.~\ref{fig:quadruplets} for
examples. Each trajectory pair $(\alpha,\gamma)$ typically
contains a huge number of encounters, where stretches of $\alpha$
and/or $\gamma$ come close to each other (up to time reversal).
Partner trajectories $\beta$, $\delta$ can be obtained by
switching connections within some encounters. Together, $\beta$
and $\delta$ go through the same links as $\alpha$ and $\gamma$,
and traverse each $l$-encounter exactly $l$ times, just like the
pair $(\alpha,\gamma)$. Consequently, the cumulative action of
$(\beta,\delta)$ is close to the one of $(\alpha,\gamma)$, with a
small action difference $\Delta
S=(S_\alpha+S_\gamma)-(S_\beta+S_\delta)$ originating from the
intra-encounter reconnections.

Different quadruplet families are distinguished by the number of
$l$-encounters, the mutual orientation of encounter stretches,
their distribution among $\alpha$ and $\gamma$, and the
reconnections leading to $\beta$ and $\delta$.  Similar as the
diagonal quadruplets, some families of quadruplets involve one
partner trajectory whose initial and final links practically
coincide with those of $\alpha$ and the other one whose initial
and final links coincide with those of $\gamma$. When these
families are depicted schematically with encounters suppressed
they all look the same and in fact like diagonal quadruplets (see
Fig.~\ref{fig:quadruplets}a), for which reason we shall refer to
them as ``$d${\it -families}''; examples are depicted in
Figs.~\ref{fig:quadruplets}b-h. In analogy to the diagonal
quadruplets, $d$-quadruplets contribute with altogether
$N_1^2N_2^2+N_1N_2$ channel combinations.  Of these, $N_1^2N_2^2$
arise when $\alpha$ and $\beta$ start and end alike since then the
four channel indices involved are unrestricted; when $\alpha$ and
$\delta$ start and end alike, $N_1N_2$ combinations arise since
the channels are restricted as $a_1=c_1, \, a_2=c_2$.

A second type of quadruplet families is drawn schematically in
Fig.~\ref{fig:quadruplets}i: here one partner trajectory
practically coincides at its beginning with $\alpha$ and at its
end with $\gamma$; the other  trajectory coincides at its
beginning with $\gamma$ and at its end with $\alpha$. The simplest
example of such an ``$x${\it -family}'' involves just one
2-encounter, see Fig.~\ref{fig:quadruplets}j
\cite{Schanz,EssenShot}; not surprisingly, our schematic sketch
strongly resembles that picture. Since quadruplets contribute to
the conductance variance only if they connect channels as
$\alpha,\beta:a_1\to a_2$ and $\gamma,\delta:c_1\to c_2$,
$x$-families arise only if either the ingoing channels or the
outgoing channels coincide. If the ingoing channels coincide,
$a_1=c_1$, the trajectory coinciding initially with $\gamma$ and
finally with $\alpha$ has the form $a_1=c_1\to a_2$ and may be
chosen as $\beta$; the trajectory coinciding initially with
$\alpha$ and finally with $\gamma$ is of the form $a_1=c_1\to c_2$
and may be chosen as $\delta$. If $a_2=c_2$, similar arguments
hold, with $\beta$ and $\delta$ interchanged. Thus, $x$-families
arise for $N_1N_2^2$ channel combinations with $a_1=c_1$, and for
$N_1^2N_2$ combinations with $a_2=c_2$, altogether for $NN_1N_2$
possibilities.

We shall presently find that quadruplet families contribute to the
conductance variance according to the same rules as do pairs to
the mean conductance: Each link yields a factor $\frac{1}{N}$,
each encounter a factor $-N$; moreover, we have to multiply with
the number of channel combinations, i.e. $N_1^2N_2^2+N_1N_2$ for
$d$-families and $NN_1N_2$ for $x$-families.

To justify these rules we consider a family with numbers of
$l$-encounters given by $\vec{v}=(v_2,v_3,v_4,\ldots)$. Again,
$\vec{v}$ determines the total number of encounters $V(\vec{v})$
and the number of encounter stretches $L(\vec{v})$.  The overall
number of links is now given by $L(\vec{v})+2$, since there is one
link preceding each of the $L(\vv)$ encounter stretches, and the
two final links of $\alpha$ and $\gamma$ which do not precede any
encounter stretch.  Similarly as for trajectory pairs, we can
determine a density $w(s,u)$ of stable and unstable separations;
this density will be normalized such that integration over all
$s,u$ belonging to an interval $(\Delta S, \Delta S+d\Delta S)$ of
action differences $\Delta S$ yields the number of pairs
$\beta,\delta$ differing from given $\alpha,\gamma$ such that the
quadruplet $(\alpha,\beta,\gamma,\delta)$ belongs to a given
family and the action difference is inside that interval. Using
the same arguments as in Subsection \ref{sec:conductance_general},
one finds $w(s,u)$ as an integral over
$\{\Omega^{L-V}\,\prod_{\sigma=1}^Vt_{\rm
  enc}^\sigma(s,u)\}^{-1}$, with the integration running over the
durations of all links, except the final links of $\alpha$ and
$\gamma$. The integration range must be restricted such that all
links (including the final ones) have positive durations. To
evaluate the contribution of one family to the quadruple sum in
(\ref{quadsum}), we may now replace the summation over $\beta$ and
$\delta$ by integration over $w(s,u)$,
\begin{equation}
\label{quadsum_w}
\big\langle(\tr(tt^\dagger))^2\big\rangle\big|_{\rm fam}
=\frac{1}{T_H^2}\left\langle\sum_{a_1,c_1\atop a_2,c_2}\int d^{L-V}s\;
d^{L-V}u\sum_{\alpha:a_1\to a_2\atop\gamma: c_1\to c_2}
|A_\alpha|^2 |A_\gamma|^2 w(s,u)
{\rm e}^{{\rm i}\sum_{\sigma,j}s_{\sigma j}u_{\sigma j}/\hbar}\right\rangle \,.
\end{equation}

The sums over $\alpha$, $\gamma$ can be performed using the
Richter/Sieber rule to ultimately get further integrals over the
durations of the final links of $\alpha$ and $\gamma$, with an
integrand involving the survival probability
$\exp\{-\frac{N}{T_H}(\sum_{i=1}^{L+2}t_i +\sum_{\sigma=1}^Vt_{\rm
  enc}^\sigma(s,u))\}$.  We thus meet with link and encounter
integrals of the same type as for the mean conductance.  All
powers of $T_H$ mutually again cancel, and we are left with a
factor $\frac{1}{N}$ from each of the $L(\vec{v})+2$ links and a
factor $-N$ from each of the $V(\vec{v})$ encounters which
altogether give
$\frac{(-1)^{V(\vec{v})}}{N^{L(\vec{v})-V(\vec{v})+2}}$.  The
summation over $a_1,c_1,a_2,c_2$ yields the number of channel
combinations mentioned.

If we denote by ${\cal N}_d(\vec{v})$, ${\cal N}_x(\vec{v})$ the
numbers of $d$- and $x$-families associated to $\vec{v}$, the sum over
all families with fixed $L(\vec{v})-V(\vec{v})=m$ involves the subsums
\begin{eqnarray}
\label{dmxm}
d_m&=&\sum_{\vec{v}}^{L(\vec{v})-V(\vec{v})=m}(-1)^{V(\vec{v})}
{\cal N}_d(\vec{v})\nonumber\\
x_m&=&\sum_{\vec{v}}^{L(\vec{v})-V(\vec{v})=m}(-1)^{V(\vec{v})}
{\cal N}_x(\vec{v})
\end{eqnarray}
which allow to write the yield of all families as
\begin{equation}
\label{variance_DX}
\big\langle(\tr(tt^\dagger))^2\big\rangle
=(N_1^2N_2^2+N_1N_2)
\underbrace{\left(\frac{1}{N^2}
+\sum_{m=1}^\infty\frac{d_m}{N^{m+2}}\right)}_{\equiv D}+NN_1N_2
\underbrace{\sum_{m=1}^\infty\frac{x_m}{N^{m+2}}}_{\equiv X}\,,
\end{equation}
with $D$ arising from $d$-families (including the diagonal
contribution) and $X$ from $x$-families.

The coefficients $d_m,x_m$ are obtained by {\it counting families of
  quadruplets.} That counting is an elementary task for small $m$.
The coefficient $d_1$ accounts for families of $d$-quadruplets
differing in one 2-encounter ($L=2$, $V=1$, $m=1$). In the unitary
case there are no such families, i.e., $d_1=0$. In the orthogonal
case, we must consider quadruplets with one partner trajectory
differing from $\alpha$ in a 2-encounter, and one partner trajectory
identical to $\gamma$, see Fig.~\ref{fig:quadruplets}b; the quadruplet
thus contains one Richter/Sieber pair and one diagonal pair. A similar
family of quadruplets involves one partner trajectory identical to
$\alpha$, and one partner trajectory differing from $\gamma$ in a
2-encounter. We thus have $d_1=-2$.

The following coefficient $d_2$ is determined by $d$-quadruplets
differing in two 2-encounters or in one 3-encounter, the latter
quadruplets contributing with a negative sign.  Some of these
quadruplets fall into two pairs contributing to the average
conductance.  Quadruplets consisting of one diagonal pair and one pair
contributing to the coefficient $c_2$ of the average conductance (see
Fig.~\ref{fig:conductance_next}) yield a contribution $2c_2$ to $d_2$
(i.e., 0 in the unitary case and 2 in the orthogonal case); the factor
2 arises because either $\alpha$ or $\gamma$ may belong to the
diagonal pair. In the orthogonal case, there is one further family of
quadruplets consisting of two Richter/Sieber pairs.  Finally, we must
reckon with quadruplets that do not fall into two pairs contributing
to the mean conductance, as depicted in
Figs.~\ref{fig:quadruplets}c-e, for the unitary case. Two further
families are obtained by ``reflection'', i.e., interchanging $\alpha$
and $\gamma$ in Figs. ~\ref{fig:quadruplets}d and
\ref{fig:quadruplets}e.  Taking into account the negative sign for
Fig.~\ref{fig:quadruplets}e and its reflected version, the respective
contributions sum up to 1.  In the orthogonal case, the additional
families in Figs.~\ref{fig:quadruplets}f-h and the reflected versions
of Fig.~\ref{fig:quadruplets}g and h yield a further summand 1.
Altogether, we thus find $d_2=1$ in the unitary case and
$d_2=2+1+1+1=5$ in the orthogonal case.

The most important family of $x$-quadruplets, see
Fig.~\ref{fig:quadruplets}j, involves a parallel encounter between one
stretch of $\alpha$ and one stretch of $\gamma$. This family,
discovered in \cite{Schanz} for quantum graphs, gives rise to a
coefficient $x_1=-1$ for systems with or without time-reversal
invariance.

With the coefficients $d_1, d_2, x_1$, the conductance variance
(\ref{variance_DX}) can be evaluated up to corrections of order
$O(\frac{1}{N})$. The result\footnote{In counting orders we assume
  that all numbers of channels are of the same order of magnitude.}
\begin{equation}
\label{variance_leading}
\big\langle(\tr(tt^\dagger))^2\big\rangle
-\big\langle\tr(tt^\dagger)\big\rangle^2
=\begin{cases}
\frac{N_1^2N_2^2}{N^4}+{\cal O}\left(\frac{1}{N}\right)
&\mbox{unitary case}\\
\frac{2N_1^2N_2^2}{N^4}+{\cal O}\left(\frac{1}{N}\right)
&\mbox{orthogonal case}
\,,
\end{cases}
\end{equation}
coincides with the random-matrix prediction (\ref{variance_RMT}).
We note that Eq.~(\ref{variance_leading}) could ultimately be
attributed only to the quadruplets shown in
Fig.~\ref{fig:quadruplets}c-h, since all other contributions
mutually cancel. (In particular, the contributions proportional to
$N_1^2N_2^2$ from all $d$-quadruplets that consist of two pairs
contributing to the conductance are cancelled by the squared
average conductance. The term proportional to $N_1N_2$ in the
diagonal approximation is compensated by the contribution of
$x$-quadruplets as in Fig.~\ref{fig:quadruplets}j.)

To go beyond Eq.~(\ref{variance_leading}) (and to show that no
terms were missed in Eq. (\ref{variance_leading})), we must
systematically count families of $d$- and $x$-quadruplets with
arbitrarily many encounters. Similar to the case of conductance
this can be done by establishing relations between families of
trajectory quadruplets and structures of periodic orbit pairs.
For details see Appendix \ref{sec:orbits}; the results differ for
the two universality classes.

In the {\it unitary case}
 we find
\begin{equation}\label{xd unitary}
d_m=\begin{cases}
0&\mbox{if $m$ odd}\\
1&\mbox{if $m$ even\,,}
\end{cases}\qquad x_m=\begin{cases}
-1&\mbox{if $m$ odd}\\
0&\mbox{if $m$ even\,.}
\end{cases}
\end{equation}
The total contributions of all $d$- and $x$-families (per channel
combination) now read
\begin{eqnarray}
\label{DX_unit}
D&=&\frac{1}{N^2}+\sum_{m+1}^{\infty}\frac{d_m}{N^{m+2}}=\frac{1}{N^2-1}\nonumber\\
X&=&\sum_{m=1}^{\infty}\frac{x_m}{N^{m+2}}=-\frac{1}{N(N^2-1)}\,.
\end{eqnarray}
The resulting conductance variance
\begin{equation}
\big\langle(\tr(tt^\dagger))^2\big\rangle-\big\langle\tr(tt^\dagger)\big\rangle^2=
\frac{N_1^2N_2^2}{N^2(N^2-1)}
\end{equation}
agrees with the random-matrix prediction (\ref{variance_RMT}).

In the {\it orthogonal case} we have
\begin{eqnarray}\label{xd orthogonal}
d_m=(-1)^{m}\frac{3^{m}+1}{2},\quad
x_m=(-1)^{m}\frac{3^{m}-1}{2}\,.
\end{eqnarray}
The contributions of $d$- and $x$-families per channel combination now
read
\begin{eqnarray}
\label{DX_orth}
D&=&\frac{1}{N^2}+\sum_{m=1}^{\infty}\frac{d_m}{N^{m+2}}=\frac{N+2}{N(N+1)(N+3)}
\,,\nonumber\\
X&=&\sum_{m=1}^{\infty}\frac{x_m}{N^{m+2}}=-\frac{1}{N(N+1)(N+3)}
\end{eqnarray}
and determine the variance in search as
\begin{equation}
\label{variance_orth}
\big\langle(\tr(tt^\dagger))^2\big\rangle-\big\langle\tr(tt^\dagger)
\big\rangle^2=\frac{2N_{1}N_{2}(N_{1}+1)(N_{2}+1)}{N(N+1)^{2}(N+3)}\,,
\end{equation}
again in agreement with (\ref{variance_RMT}).  Thus, we have once more
verified the universal behavior of individual chaotic cavities.

\section{Shot noise}

Our reasoning can be extended to a huge class of observables which are
quartic in the transmission amplitudes and thus determined by $d$- and
$x$-quadruplets as well. For a first example, we consider shot noise: Due to
the discreteness of the elementary charge, the current flowing through
a mesoscopic cavity fluctuates in time as $I(t) = \overline{I}+\delta
I(t)$ where $\overline{I}$ denotes the average current.  These current
fluctuations, the so-called shot noise, remain in place even at zero
temperature. They are usually characterized through the power
\cite{Beenakker}

\begin{equation}
P=4\int_0^\infty\overline{\delta I(t_0)\delta I(t_0+t)}dt
\end{equation}
where the overline indicates an average over the reference time
$t_0$.\footnote{ This definition, as well as the treatment of
three-lead correlation in the following section, follows the
conventions of \cite{Beenakker,ThreeLeadL}, and differs by a
factor 2 from \cite{ThreeLeadB}.}

Using our semiclassical techniques, we proceed to showing that for
chaotic cavities, the energy-averaged power of shot noise takes a
universal form. Again, our treatment applies to {\it individual}
cavities and yields an expansion to {\it all orders in the inverse
  number of channels}. That expansion turns out convergent and
summable to a simple expression which subsequently to our prediction
was checked to agree with random-matrix theory by Savin and Sommers
\cite{Dima}.

Following B{\"u}ttiker
\cite{Buettiker}, we express the power of shot noise through the
transition matrices $t$
\begin{equation}
\langle P\rangle=\langle\tr(tt^\dagger)-\tr(tt^\dagger tt^\dagger)
\rangle\,;
\end{equation}
here, $P$ is averaged over the energy and measured in units
$\frac{2e^3|V|}{\pi\hbar}$ depending on the voltage $V$.  While the
average conductance $\langle\tr(tt^\dagger)\rangle$ was already
evaluated in Section \ref{sec:conductance}, the quartic term turns
into a quadruple sum over trajectories similar to the conductance
variance
\begin{eqnarray}
\big\langle(\tr(tt^\dagger tt^\dagger))\big\rangle
=\left\langle\sum_{a_1,c_1\atop a_2,c_2}t_{a_1a_2}t_{c_1a_2}^*
t_{c_1c_2}t_{a_1c_2}^*\right\rangle
=\frac{1}{T_H^2}\left\langle\sum_{a_1,c_1\atop a_2,c_2}\;
\sum_{\substack{
\alpha:\,a_1\to a_2\\
\beta:\, c_1 \to a_2 \\
\gamma:\, c_1\to c_2\\
\delta:\,a_1 \to c_2}}
A_\alpha A_\beta^*A_\gamma A_\delta^*
{\rm e}^{{\rm i}(S_\alpha-S_\beta+S_\gamma-S_\delta)/\hbar}
\right\rangle\,;
\end{eqnarray}
the trajectories $\alpha$, $\beta$, $\gamma$, $\delta$ must now
connect the ingoing channels $a_1,c_1$ to the outgoing channels
$a_2, c_2$ as indicated in the summation prescription.

As a
consequence, the possible {\it channel combinations} for $d$- and
$x$-families of quadruplets are changed relative to the
conductance variance. In the present case, {\it $d$-quadruplets},
with one partner trajectory coinciding at its beginning and end
with $\alpha$, and the other partner trajectory doing the same
with $\gamma$, contribute only if either the ingoing or the
outgoing channels coincide.  If the ingoing channels coincide, the
partner trajectory connecting the same points as $\alpha$ is of
the type $a_1 = c_1 \to a_2$ and may be taken as $\beta$, whereas
the trajectory connecting the same points as $\gamma$ has the form
$a_1 = c_1 \to c_2$ and may be chosen as $\delta$.  If the
outgoing channels coincide, similar arguments apply, with $\beta$
and $\delta$ interchanged. Thus, $d$-quadruplets contribute only
for $N_1N_2^2 + N_1^2N_2 = NN_1N_2$ channel combinations. In this
sense, they take the role played by $x$-quadruplets in case of the
conductance variance.

In turn, {\it $x$-quadruplets} now contribute for all channel
combinations. Moreover, if both the ingoing and the outgoing channels
coincide, either of the two partner trajectories may be chosen as
$\beta$ or $\delta$, meaning that the corresponding channel
combinations have to be counted for a second time.  Thus,
$x$-quadruplets now contribute for altogether $N_1^2N_2^2+N_1N_2$
channel combinations, like $d$-quadruplets in case of the conductance
variance.

\begin{figure}
\begin{center}
\includegraphics[scale=0.9]{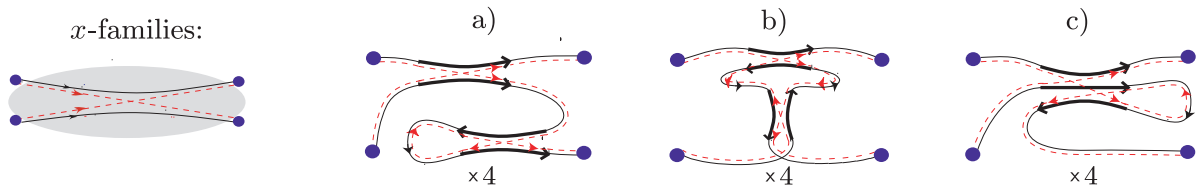}
\end{center}
\caption{Families of $x$-quadruplets with $m=L-V=2$,
  contributing to the next-to-leading order of shot noise for
  time-reversal invariant systems.}
\label{fig:shot}
\end{figure}

We can simply interchange the multiplicity factors in our formula for
$\langle(\tr(tt^\dagger))^2\rangle$, Eq. (\ref{variance_DX}), to get
\begin{equation}
\label{shot_DX} \big\langle\tr(tt^\dagger
tt^\dagger)\big\rangle= NN_1N_2D+(N_1^2N_2^2+N_1N_2)X
\end{equation}
and thus
\begin{equation}\label{univshot}
\langle P\rangle=
\begin{cases} \frac{N_1^2 N_2^2}{N(N^2-1)} &\text{unitary case}\\
  \frac{N_1(N_1+1)N_2(N_2+1)}{N(N+1)(N+3)}&\text{orthogonal case}\,.
\end{cases}
\end{equation}
Eq. (\ref{univshot}) extends the known random-matrix result
\cite{Beenakker},
\begin{equation}  \label{shot_RMT}
\langle P\rangle=\begin{cases}
\frac{N_1^2N_2^2}{N^3}+{\cal O}\Big(\frac{1}{N}\Big)&\text{unitary case}\\
\frac{N_1^2N_2^2}{N^3}+\frac{N_1N_2(N_1-N_2)^2}{N^4}+{\cal O}\Big(\frac{1}{N}\Big)&\text{orthogonal case}\,,
\end{cases}
\end{equation}
to all orders in $\frac{1}{N}$, for individual chaotic cavities.
We can, moreover, give an intuitive {\it interpretation} for the
terms in (\ref{shot_RMT}). The diagonal contributions to
$\langle\tr(tt^\dagger)\rangle$ and $\langle\tr(tt^\dagger
tt^\dagger)\rangle$ both read $\frac{N_1N_2}{N}$ and therefore
mutually cancel. The leading contribution,
$\frac{N_1^2N_2^2}{N^3}$, arises from $d$-quadruplets differing in
a single 2-encounter (see Fig. \ref{fig:quadruplets}j). In the
unitary case, there are no terms of order 1, since all related
families require time-reversal invariance. In the orthogonal case,
Richter/Sieber pairs yield a contribution $-\frac{N_1 N_2}{N^2}$
to $\langle\tr(tt^\dagger)\rangle$, from which we have to subtract
two contributions to $\langle\tr(tt^\dagger tt^\dagger)\rangle$,
the term $-\frac{2N_1N_2}{N^2}$ accounting for $d$-quadruplets
differing in a single antiparallel 2-encounter (see Fig.
\ref{fig:quadruplets}b), and a term $\frac{4N_1^2N_2^2}{N^4}$
arising from $x$-quadruplets contributing to $x_2=4$. The latter
$x$-quadruplets may differ in two 2-encounters, as in Figs.
\ref{fig:shot}a and \ref{fig:shot}b, or in one 3-encounter, as in
Fig. \ref{fig:shot}c. From the examples in Fig. \ref{fig:shot},
further families are obtained by interchanging $\alpha$ and
$\gamma$, interchanging the two leads (for Figs. \ref{fig:shot}a
and \ref{fig:shot}c), or interchanging the pairs $(\alpha,\gamma)$
and $(\beta,\delta)$ (for Fig. \ref{fig:shot}b). Each of the Figs.
\ref{fig:shot}a-c therefore represents altogether four families,
whose contributions indeed sum up to $x_2=4(-1)^2+4(-1)^2+4(-1)=4$.
Together with the contributions mentioned before, they combine to
(\ref{shot_RMT}).\footnote{In \cite{Whitney}, trajectory
quadruplets
  where the encounter directly touches the lead are shown to become
  relevant when the mean dwell time is of the order of the Ehrenfest
  time.}

\section{Current correlations in cavities with three leads}

\begin{figure}
\begin{center}
\includegraphics[scale=0.8]{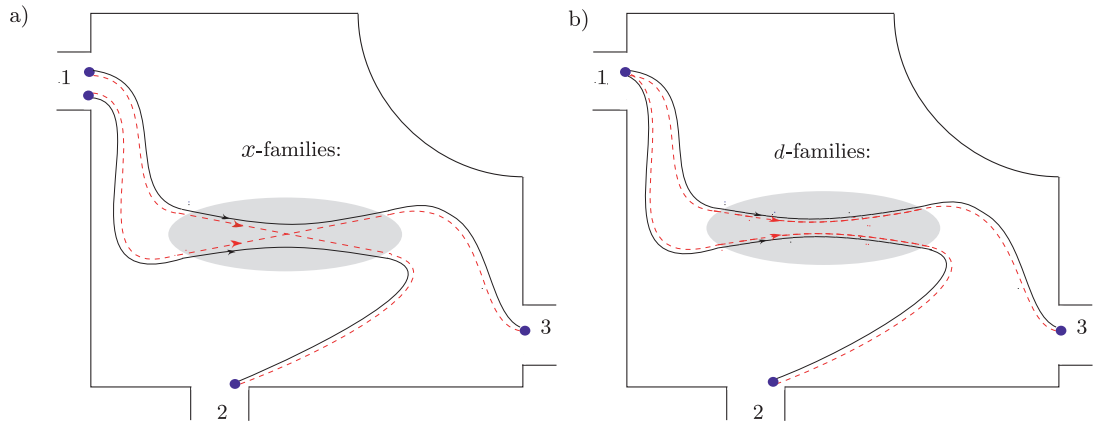}
\end{center}
\caption{
  A cavity with three leads. If a voltage $V$ is applied between lead
  1 and leads 2 and 3, one observes currents $I^{(1\to 2)}$ and
  $I^{(1\to 3)}$.  As explained in the text, correlations between these
  currents are again determined by families of $d$- and
  $x$-quadruplets of trajectories.}
\label{fig:threelead}
\end{figure}

Another interesting experimental setting involves a chaotic cavity
with {\it three leads}, respectively supporting $N_1$, $N_2$ and
$N_3$ channels; see Fig.~\ref{fig:threelead}. The second and the
third lead are kept at the same potential, and a voltage is
applied between these leads and the first one. Consequently,
currents $I^{(1\to 2)}$, $I^{(1\to 3)}$ flow from the first lead
to the second and third one. We shall be interested in the
fluctuations $\delta I^{(1\to 2)}$, $\delta I^{(1\to 3)}$ of these
currents around the corresponding averages values, and study
correlations between $\delta I^{(1\to 2)}$ and $\delta I^{(1\to
3)}$ \cite{ThreeLeadL,ThreeLeadB}.

 This setting is similar to the famous {\it
  Hanbury Brown-Twiss experiment} \cite{HBT} in quantum optics: there,
light from some source (corresponding to the first lead) was
detected by two photomultipliers (corresponding to the second and
third lead). Similar work on Fermions began somewhat later
\cite{ThreeLeadL,ThreeLeadB} but the precise from of the
correlation function is as yet unknown.

Our semiclassical reasoning can easily be extended to fill this
gap. The two currents depend on the matrices $t^{(1\to 2)}$,
$t^{(1\to3)}$ containing the transition amplitudes between
channels of the first and the second and third lead; these
matrices have the sizes $N_1 \times N_2$ and $N_1 \times N_3$. As
shown in \cite{ThreeLeadL,ThreeLeadB},
correlations between $\delta
I^{(1\to 2)}$ and $\delta I^{(1\to 3)}$
are determined by the transition amplitudes as
\begin{equation}
4\int_0^\infty \overline{\delta I^{(1\to2)}(t_0) \delta I^{(1\to
3)}(t_0 + t)} d t
 = -  \big\langle
{\rm tr} (t^{(1\to 2)}{t^{(1\to 2)}}^\dagger t^{(1\to 3)}{t^{(1\to
3)}}^\dagger)\big\rangle\
\end{equation}
(in units of $\frac{2e^3 |V|}{\pi\hbar}$). Using the semiclassical
expression for the transition amplitudes, we are again led to a
sum over quadruplets of trajectories
\begin{eqnarray}
\big\langle\tr (t^{(1\to 2)}{t^{(1\to 2)}}^\dagger
t^{(1\to 3)}{t^{(1\to 3)}}^\dagger)\big\rangle
=\left\langle
\sum_{\substack{
a_1,c_1=1\ldots N_1\\ \quad\;
a_2=1\ldots N_2,\\\quad\;
c_3=1\ldots N_3}}
t^{(1\to 2)}_{a_1a_2}{t^{(1\to 2)}_{c_1a_2}}^*t^{(1\to 3)}_{c_1c_3}
{t^{(1\to 3)}_{a_1c_3}}^*\right\rangle\nonumber\\
=\frac{1}{T_H^2}\left\langle
\sum_{a_1,c_1,a_2,c_3}\;
\sum_{\substack{
\alpha: \,a_1\to a_2\\
\beta: \,c_1\to a_2\\
\gamma:\, c_1\to c_3\\
\delta:\, a_1\to c_3}}
A_\alpha A_\beta^*A_\gamma A_\delta^*
{\rm e}^{{\rm i}(S_\alpha-S_\beta+S_\gamma-S_\delta)/\hbar}
\right\rangle\,,
\end{eqnarray}
with $a_1$, $c_1$, $a_2$, $c_3$ labelling channels of the first,
second and third lead, as indicated by the subscript. The
trajectories $\alpha$, $\beta$, $\gamma$, $\delta$ must connect
these channels as $\alpha(a_1\to a_2)$, $\beta(c_1\to a_2)$,
$\gamma(c_1\to c_3)$, $\delta(a_1\to c_3)$.

The contribution of each family of trajectory quadruplets can be
evaluated similarly to the conductance variance or shot noise.
Since a particle can leave the cavity through any of the three
leads, the {\it escape rate} depends on the overall number of
channels $N = N_1 + N_2 + N_3$. Again, integration brings about
factors $\frac{1}{N}$ and $-N$ for each link and each encounter.
Only the numbers of channel combinations are changed. {\it
$x$-quadruplets} as in Fig. \ref{fig:threelead}a contribute for
all $N_1^2N_2N_3$ possible choices of $a_1$, $c_1$, $a_2$, $c_3$.
For any of these choices, partner trajectories connecting the
initial point of $\gamma(c_1\to c_3)$ to the final point of
$\alpha(a_1\to a_2)$, and the initial point of $\alpha$ to the
final point of $\gamma$ are of the form $c_1\to a_2$ and $a_1\to
c_3$ and can be chosen as $\beta$ and $\delta$, respectively.  In
contrast, {\it
  $d$-quadruplets} as in Fig.~\ref{fig:threelead}b contribute only
  for
the $N_1N_2N_3$ combinations with coinciding ingoing channels $a_1 =
c_1$. For these combinations the partner trajectory coinciding at its
ends with $\alpha$ is of the type $a_1 = c_1 \to a_2$ and can be taken as
$\beta$ whereas the partner trajectory coinciding at its ends with $\gamma$
has the form $a_1 = c_1 \to c_3$ and can be chosen as $\delta$. With these
numbers of channel combinations, the current correlations in a 3-lead
geometry are obtained as
\begin{equation}
\langle \tr(t^{(1\to 2)}{t^{(1\to 2)}}^\dagger
t^{(1\to 3)}{t^{(1\to 3)}}^\dagger)\rangle
=N_1N_2N_3D+N_1^2N_2N_3X=\begin{cases}
\frac{N_1N_2N_3(N_2+N_3)}{N(N^2-1)}&\text{unitary case}\\
\frac{N_1N_2N_3(N_2+N_3+2)}{N(N+1)(N+3)}&\text{orthogonal case}\,.
\end{cases}
\end{equation}

\section{Ericson fluctuations}

\label{sec:ericson}

Another interesting quantum signature of chaos are so-called
Ericson fluctuations, which have first been discovered
experimentally in nuclear physics. In compound-nucleus reactions
with strongly overlapping resonances, universal fluctuations in
the correlation of two scattering cross sections at different
energies have been observed. A first interpretation in terms of
random-matrix theory was provided by Ericson \cite{Ericson} and
further theoretical investigations have been reported in
\cite{VWZ}, \cite{DB}.

Later on, the relation between classical chaotic scattering and
Ericson fluctuations in single-particle quantum mechanics has been
discussed \cite{BS}. Theoretical work shows that, e.g. the
photoionization cross section of Rydberg atoms in external fields
show universal correlations once the underlying classical dynamics
is chaotic \cite{MainWunner}.

Chaotic transport through a ballistic cavity displays
Ericson fluctuations in the covariance of the conductance at
{\it two different energies},

\begin{equation}
\big\langle C(E,\epsilon)\big\rangle=\left\langle G(E)G
\left(E+\frac{\epsilon
N}{2\pi\overline{\rho}}\right)\right\rangle
-\langle G\rangle^2\,,
\end{equation}
with $G(E) = \tr(tt^\dagger)$. Here, the difference between the
two energies was made dimensionless by referral to the energy
scale $\frac{N}{2\pi\overline{\rho}}$ proportional to the number
of channels and to the mean level spacing. Similarly as for the
conductance variance, the semiclassical approximation
(\ref{transition}) for $t(E),t(E')$ where $E'=E+\frac{\epsilon
  N}{2\pi\overline{\rho}}$ leads to a
quadruple sum over trajectories,
\begin{eqnarray}
\label{quadsum_covariance} \big\langle
\tr\;tt^\dagger(E)\;\;\tr\;tt^\dagger(E')\big\rangle
&=&\left\langle\sum_{a_1,c_1\atop
a_2,c_2}t_{a_1a_2}(E)t_{a_1a_2}^*(E)
t_{c_1c_2}(E')t_{c_1c_2}^*(E')\right\rangle\\
&=&\frac{1}{T_H^2}\left\langle \sum_{a_1,c_1\atop
a_2,c_2}\;\sum_{\substack{
\alpha,\beta:\, a_1\to a_2\\
\gamma, \delta:\,c_1\to c_2}} A_\alpha A_\beta^*A_\gamma
A_\delta^* {\rm
e}^{{\rm i}(S_\alpha(E)-S_\beta(E)+S_\gamma(E')-S_\delta(E' ))/\hbar}
\right\rangle\,,\nonumber
\end{eqnarray}
the only difference to (\ref{quadsum}) being that the trajectories
$\gamma$ and $\delta$ have to be taken at energy
$E'=E+\frac{\epsilon
  N}{2\pi\overline{\rho}}$.  Using  $\frac{\partial
  S_\gamma}{\partial E}=T_\gamma$, $\frac{\partial S_\delta}{\partial
  E}=T_\delta$, and $T_H=2\pi\hbar\overline{\rho}$, the phase factor
can be cast into the form
\begin{equation}
{\rm e}^{{\rm i}(S_\alpha(E)-S_\beta(E)
+S_\gamma(E+\frac{\epsilon N}{2\pi\overline{\rho}})
-S_\delta(E+\frac{\epsilon N}{2\pi\overline{\rho}}))/\hbar}
\approx
{\rm e}^{{\rm i}(S_\alpha(E)-S_\beta(E)+S_\gamma(E)-S_\delta(E))/\hbar}
\times{\rm e}^{{\rm i}\frac{N}{T_H}\epsilon(T_\gamma-T_\delta)}\,,
\end{equation}
i.e., the quadruple sum in (\ref{quadsum_covariance}) differs from
(\ref{quadsum}) by an additional factor depending on the {\it
  difference between the dwell times of $\gamma$ and $\delta$}.

The latter difference may be written as a sum over links (with
durations $t_i$) and encounters (with durations $t_{\rm
enc}^\sigma$),
\begin{equation}
T_\gamma-T_\delta=\sum_{i=1}^{L+2}
\eta_i t_i+\sum_{\sigma=1}^V\eta_{\sigma}t_{\rm enc}^{\sigma}\,.
\end{equation}
Here, the integer numbers $\eta_i$ and $\eta_{\sigma}$
characterize the individual links and encounters.  (Note the
distinction between links and encounters by Latin and Greek
subscripts). Each link occurs twice in the quadruplet, once in one
of the original trajectories $\alpha,\gamma$ and then in one of
the partner trajectories $\beta,\delta$. The number
$\eta_i=0,\pm1$ gives the difference between the numbers of times
the $i$-th link is traversed by the trajectories $\gamma$ and
$\delta$. We thus have $\eta_i = 1$ if the $i$-th link is
traversed by $\gamma$ and not by $\delta$, $i = 0$ if it is
traversed either by both or none of the two trajectories, and $i =
-1$ if it is traversed only by $\delta$.  Similarly,
$\eta_{\sigma}$ gives the difference between the numbers of
traversals of the $\sigma$-th encounter by $\gamma$ and $\delta$.
For an $l$-encounter, $\eta_{\sigma}$ may range between $-l$ and
$l$.

When evaluating the contribution of each family of quadruplets, Eq.
(\ref{quadsum_w}), we simply have to add a phase factor ${\rm e}^{{\rm
    i}\frac{N}{T_H}\eta_i t_i}$ for each link, and a phase factor
${\rm e}^{{\rm i}\frac{N}{T_H}\eta_{\sigma} t_{\rm enc}^{\sigma}}$
for each encounter. The link and encounter integrals are thus
replaced by
\begin{equation}
\int_0^\infty dt_i{\rm e}^{-\frac{N}{T_H}t_i}
{\rm e}^{{\rm i}\frac{N}{T_H}\eta_i t_i}
=\frac{T_H}{N(1-{\rm i}\eta_i\epsilon)}\,,
\end{equation}
and
\begin{equation}
\left\langle\int d^{l-1}s d^{l-1}u
\frac{1}{\Omega^{l-1}\,t_{\rm enc}^{\sigma}(s,u)}
{\rm e}^{-\frac{N}{T_H}t_{\rm enc}(s,u)}
{\rm e}^{{\rm i}\frac{N}{T_H}\eta_{\sigma} t_{\rm enc}(s,u)^{\sigma}}
{\rm e}^{{\rm i}\sum_j s_{\sigma j}u_{\sigma j}/\hbar}\right\rangle
=-\frac{N(1-{\rm i}\eta_{\sigma}\epsilon)}{T_H^{l-1}}\,.
\end{equation}
As a consequence, our diagrammatic rules are modified to yield {\it a
  factor $\frac{1}{N(1-{\rm i}\eta_i\epsilon)}$ for each link and a factor
  $-N(1-{\rm i}\eta_{\sigma}\epsilon)$ for each encounter}.

We must, however, be aware that the numbers $\eta_i,\eta_{\sigma}$
depend on which of the two partner trajectories is labelled as $\beta$
and which is labelled as $\delta$. Each family of $d$- or
$x$-quadruplets hence comes with {\it two different sets of numbers
  $\{\eta_i,\eta_{\sigma}\}$}, depending on the combinations of
channels considered.  For each $d$-family we have to keep into
account $N_1^2N_2^2$ channel combinations with $\delta$ coinciding
at its ends with $\gamma$; all these ``$\delta\gamma$-type''
combinations give rise to the same $\{\eta_i,\eta_{\sigma}\}$ and
to the same link and encounter factors. In addition, we must
consider $N_1N_2$ combinations of the ``$\delta\alpha$-type'' with
$\delta$ coinciding at its ends with $\alpha$, and a different set
of $\{\eta_i,\eta_{\sigma}\}$.

For each $x$-family we would, in principle, have to distinguish
between $N_1N_2^2$ combinations with coinciding ingoing channels,
and $\delta$ coinciding at its beginning  with $\alpha$ and at its
end with $\gamma$, and $N_1^2N_2$ combinations with coinciding
outgoing channels, and $\delta$ coinciding at its beginning with
$\gamma$ and at its end with $\alpha$. Such caution is, however,
unnecessary for reasons of symmetry. Each $x$-family is
accompanied by another one which is topologically
mirror-symmetrical, with left and right in
Fig.~\ref{fig:quadruplets} interchanged. In this family, initial
points turn into final ones, and vice versa, implying that $\beta$
and $\delta$ are interchanged. Since both families are taken into
account simultaneously, ``mistakes'' like always choosing $\delta$
to connect the initial point of $\alpha$ to the final point of
$\gamma$, are automatically compensated.

We can thus write the conductance
covariance as
\begin{eqnarray}
\label{Ericson genformula}
C(\epsilon)=N_1^2N_2^2\sum_{m=0}^\infty
\frac{d_m^{\;(\delta\gamma)}}{N^{m+2}}+N_1N_2\sum_{m=0}^\infty
\frac{d_m^{\;(\delta\alpha)}} {N^{m+2}}+N_1N_2N\sum_{m=1}^\infty
\frac{x_m}{N^{m+2}}-\langle G\rangle^2
\end{eqnarray}
Here the coefficients $x_m(\epsilon),d_m^{\;(\delta\gamma)}(\epsilon), d_m^{\;(\delta\alpha)}(\epsilon)$
are the summary contributions of the $x$-quadruplets and the two
mentioned groups of $d$-quadruplets, with $m=L(\vec v)-V(\vec v)-2$
(and thus $m=0$ for the diagonal quadruplets) and the denominator
$N^{-m-2}$ dropped. The squared averaged conductance is
$\epsilon$-independent and is determined by Eq.  (\ref{RMT_conductance}).

\begin{figure}
\begin{center}
\includegraphics[scale=0.9]{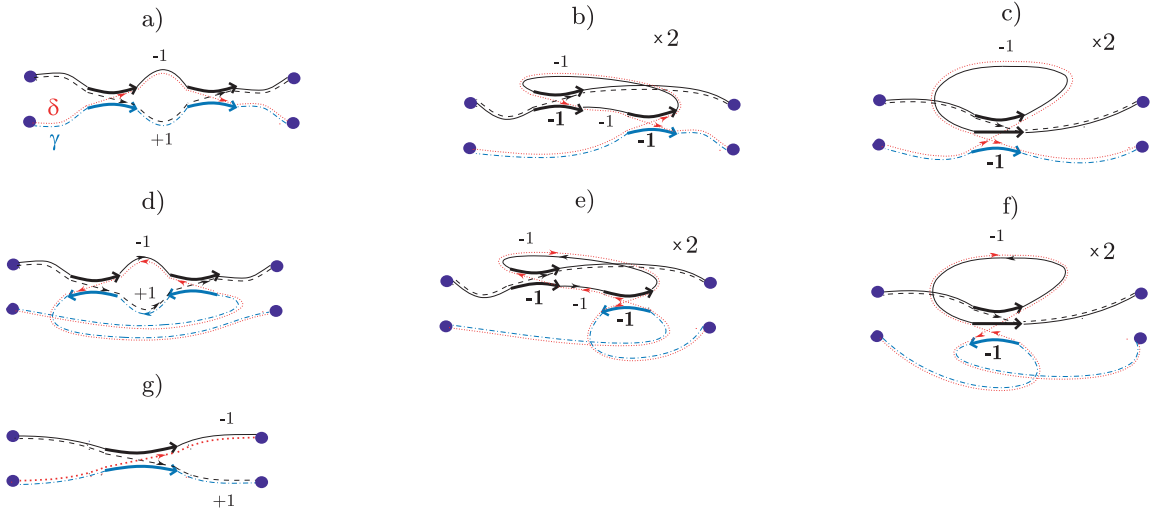}
\end{center}
\caption{ Families of trajectory quadruplets contributing to
  the covariance of conductance (coinciding with Fig.~
  \ref{fig:quadruplets}c-h,j). The trajectories $\gamma$ and $\delta$
  are highlighted through dashing and dotting, assuming that $\delta$
  connects the same points as $\gamma$. The picture also indicates all
  non-vanishing numbers $\eta_i$ and $\eta_{\sigma}$ (the latter in bold font).}
\label{fig:covariance}
\end{figure}

The {\it leading contribution to the conductance covariance}
corresponds to dropping in (\ref{Ericson genformula}) all
coefficients but
$x_1,d_0^{\;(\delta\gamma)},d_0^{\;(\delta\alpha)},d_1^{\;(\delta\gamma)},
d_2^{\;(\delta\gamma)}$.  For the conductance variance, we had
seen that the contributions of $d$-quadruplets that fall into
pairs ($\alpha$, $\beta$) and $(\gamma$, $\delta$) contributing to
conductance cancel with the squared average conductance. The same
remains valid here, since for these pairs $\gamma$ and $\delta$
traverse the same links and encounters, and all $\eta_i$ and
$\eta_{\sigma}$ vanish. Again, the contributions of diagonal
quadruplets $\alpha=\delta$, $\beta= \gamma$ and $x$-quadruplets
as in Fig.~\ref{fig:quadruplets}j  (see also
Fig.~\ref{fig:covariance}g) mutually compensate; compared to the
variance both kinds of quadruplets receive the same additional
factors due to one link with $\eta_i = 1$ (the trajectory $\beta =
\gamma$ and the lower left link in Fig.~\ref{fig:covariance}g) and
one link with $\eta_i = -1$ (the trajectory $\alpha = \delta$ and
the upper left link in Fig.~\ref{fig:covariance}g).

Like for the conductance variance, the leading contribution to the
covariance thus originates from the $d$-families in
Fig.~\ref{fig:quadruplets}c-h which contain encounters {\it
between} $\alpha$ and $\gamma$ and thus do not fall into pairs
relevant for conductance. These families are redrawn in
Fig.~\ref{fig:covariance}a-f, together with all non-vanishing
numbers $\eta_i$ and $\eta_{\sigma}$. The trajectories $\gamma$
and $\delta$ are highlighted through dashing and dotting, assuming
that $\delta$ connects the same points as $\gamma$; channel
combinations with $\delta$ connecting the same points as $\alpha$
only contribute to higher orders in $\frac{1}{N}$. The family of
Fig.~\ref{fig:covariance}a involves a link with $\eta_i = -1$ (the
upper central one) and a link with $\eta_i = 1$ (immediately
below), and thus yields $\frac{N_1^2N_2^2(-N)^2}{N^6(1+{\rm
i}\epsilon)(1-{\rm i}\epsilon)}
=\frac{N_1^2N_2^2}{N^4(1+\epsilon^2)}$. The same holds for the
family in Fig.~\ref{fig:covariance}d, which requires time-reversal
invariance. In contrast, the contributions of
Fig.~\ref{fig:covariance}b,c,e, and f remain independent of
$\epsilon$, since additional factors from links with $\eta_i = -1$
and encounters with $\eta_{\sigma} = -1$ mutually compensate; the
same applies for the families represented by ``$\times 2$'' in
Fig.~\ref{fig:covariance}, with ($\alpha$, $\beta$) and ($\gamma$,
$\delta$) interchanged and the signs of $\eta_i$ and
$\eta_{\sigma}$ flipped. As for the variance of conductance, the
contributions of Fig.~\ref{fig:covariance}b,c,e,f thus mutually
cancel, both in the orthogonal case and in the unitary case (where
only Figs.~\ref{fig:covariance}b,c may exist). Altogether, we now
obtain
\begin{equation}
\label{covariance_lead}
\left\langle G(E)G\left(E+\frac{\epsilon N}{2\pi
\overline{\rho}}\right)\right\rangle
=\begin{cases}
\frac{N_1^2 N_2^2}{N^4(1+\epsilon^2)}+{\cal O}
\left(\frac{1}{N}\right) & \text{unitary case} \\
\frac{2N_1^2 N_2^2}{N^4(1+\epsilon^2)}+{\cal O}\left(\frac{1}{N}
\right) & \text{orthogonal case}\,.
\end{cases}
\end{equation}
The Lorentzian form of (\ref{covariance_lead}) confirms the
random-matrix predictions of \cite{Ericson}.

{\it Higher orders} in $\frac{1}{N}$, not known from random-matrix
theory, can be accessed by straightforward computer-assisted
counting of families of quadruples differing in a larger number of
encounters, or in encounters with more stretches. To do so, we
generated permutations which describe possible structures of orbit
pairs 
(see Appendix \ref{sec:numerics} and \cite{EssenFF}). We then ``cut'' through these 
pairs as
described in Appendices \ref{sec:orbits} and \ref{sec:numerics}
to obtain quadruplets of trajectories
and determined the
corresponding $\eta_i$, $\eta_\sigma$. The final result can be
written as
\begin{equation}
\label{covariance_numerics} \left\langle
G(E)G\left(E+\frac{\epsilon
N}{2\pi\overline{\rho}}\right)\right\rangle
=\begin{cases}\frac{N_1^2N_2^2}{N^4}
\left\{\frac{1}{(1+\epsilon^2)}
+\frac{1+3\epsilon^2+21\epsilon^4+5\epsilon^6+2\epsilon^8}
{N^2\left(1+\epsilon^2\right)^5}\right\}\\
+{\cal O}\left(\frac{1}{N^4}\right) & \text{unitary case;} \\
\frac{2N_1^2N_2^2}{N^4\left(1+\epsilon^2\right)} +
\frac{2N_1N_2}{N^3\left(1+\epsilon^2\right)}
-\frac{2N_1^2N_2^2\left(5+12\epsilon^2
+3\epsilon^4\right)}{N^5\left(1+\epsilon^2\right)^3}\\
-\frac{4N_1N_2} {N^4\left(1+\epsilon^2\right)^3}
+\frac{2N_1^2N_2^2\left(18+78\epsilon^2
+177\epsilon^4+48\epsilon^6+11\epsilon^8\right)}
{N^6\left(1+\epsilon^2\right)^5}
\\
+{\cal O}\left(\frac{1}{N^3}\right)& \text{orthogonal case}\,.
\end{cases}
\end{equation}
In the unitary case the $x$-type contribution cancels in all
orders with the $d^{\;(\delta\alpha)}$-contribution; for that reason
the overall result is proportional to $N_1^2N_2^2$.

\section{Quantum transport in the presence of a weak magnetic field}

\subsection{Changed diagrammatic rules}

Our methods can also be applied to the case of a {\it weak magnetic
  field, with a magnetic action of the order of $\hbar$}.  The
necessary modifications were introduced in \cite{Nagao} for the
spectral form factor; see also \cite{JapanEssen}.  As in \cite{Nagao}, we will obtain results
interpolating between the orthogonal case (without a magnetic field)
and the unitary case, where the magnetic field is strong enough to
fully break time-reversal invariance. We shall assume that the field
is too weak to influence the classical motion, meaning that we have to
deal with the same families of trajectory pairs as in the orthogonal
case. However, the action of each trajectory is increased by an amount
proportional to the integral of the vector potential ${\bf A}$ along
that trajectory, e.g., by
\begin{equation}
\Theta_{\alpha}=\int_{\alpha}\frac{e}{c}{\bf A}({\bf q})\cdot d{\bf q}\,,
\end{equation}
for the trajectory $\alpha$. When we evaluate the average
conductance, the action difference inside each pair of
trajectories $\alpha$ and $\beta$ is thus increased by
$\Theta_{\alpha}-\Theta_{\beta}$. This additional term may be
neglected for pairs of trajectories where all encounters are
parallel. For these pairs, all links and stretches of $\beta$ are
close in phase space to links and stretches of $\alpha$, and
therefore receive almost the same magnetic action.

The situation is different for pairs where $\alpha$ and $\beta$
traverse links or stretches with opposite sense of motion. Since
the magnetic action changes sign under time reversal, such orbit
pairs have significant magnetic action differences
$\Theta_\alpha-\Theta_\beta$.  These differences can be split into
contributions from the individual links and encounters.  Let us
first consider {\it links}.  If $\beta$ contains the time-reversed
of the $i$-th link of $\alpha$, it must obtain the negative of the
corresponding magnetic action $\Theta_i$. The difference
$\Theta_\alpha-\Theta_\beta$ then receives a contribution
$2\Theta_i$. Therefore we may write the contribution of  each link
as $2\mu_i\Theta_i$ with $\mu_i= 1$ if the link changes direction
on $\beta$ and $\mu_i = 0$ otherwise.

Consider now the contribution of {\it encounters}. We assume that in the original
trajectory $\alpha$ the encounter $\sigma$ had $\nu_\sigma$
stretches traversed in some direction (arbitrarily chosen
as``positive'') meaning that the remaining $l_\sigma-\nu_\sigma$
stretches were traversed in the opposite, ``negative'' direction;
in the trajectory partner $\beta$ these numbers will generally
change to $\nu_\sigma',l_\sigma-\nu_\sigma'$ correspondingly.
Denoting the magnetic action accumulated on a single stretch
traversed in a positive direction by $\Theta_{\sigma}$ we see that
the encounter $\sigma$ yields $2\mu_{\sigma}\Theta_{\sigma}$ to
the magnetic action difference, with
$\mu_\sigma=\nu_\sigma'-\nu_\sigma$. The overall magnetic
contribution to the action difference now reads
\begin{equation}
\Theta_{\alpha}-\Theta_{\beta}=\sum_{i=1}^{L+1}2\mu_i \Theta_i +
\sum_{\sigma=1}^V 2\mu_{\sigma}\Theta_{\sigma}
\end{equation}
and yields a phase factor
\begin{equation}
\prod_{i=1}^{L+1}{\rm e}^{{\rm i}2\mu_i\Theta_i/\hbar}\prod_{\sigma=1}^{V}{\rm e}^{{\rm i}2\mu_{\sigma}\Theta_{\sigma}/\hbar}\,,
\end{equation}
where we again distinguish between links and encounters only
through Latin vs. Greek subscripts.

To handle this additional phase factor, we show that for fully
chaotic (in particular, ergodic and mixing) dynamics, the magnetic
action may effectively be seen as a {\it random variable}
\cite{Nagao}.  For fully chaotic systems, any point on any
trajectory can be located everywhere on the energy shell, with a
uniform probability given by the Liouville measure.  Moreover,
phase-space points following each other after times larger than a
certain classical ``equilibration'' time $t_{\rm cl}$ can be seen
as uncorrelated. We will therefore split each link or encounter
stretch into pieces of duration $t_{\rm cl}$. These pieces have
different magnetic actions.  Let us consider the probability
density for these actions.  Since positive and negative
contributions to the magnetic action are equally likely, the
expectation value for the action of an orbit piece must be equal
to zero.  The width $W$ (i.e., the square root of the variance
$W^2$) must be proportional to the vector potential and therefore
to the magnetic field $B$. Since the magnetic actions of the
individual pieces are uncorrelated, the central limit theorem then
implies that the magnetic actions of links with
$K\equiv\frac{t_i}{t_{\rm cl}}\gg 1$ pieces obey a Gaussian
probability distribution with the width $\sqrt{K}W$, i.e.,
\begin{equation}
P(\Theta_i)=\frac{1}{\sqrt{2\pi KW^2}}{\rm e}^{-\frac{\Theta_i^2}{2KW^2}}
\end{equation}
The phase factor arising from a link averages to $\int d
\Theta_iP(\Theta_i){\rm e}^{{\rm i}2\mu_i\Theta_i/\hbar} ={\rm
  e}^{-\mu_i b t_i}$, depending on the system-specific parameter
$b=\frac{2KW^2}{\hbar^2 t_i}=\frac{2W^2}{\hbar^2t_{\rm
    cl}}\propto \frac{B^2}{\hbar^2}$ and on $\mu_i=\mu_i^2\in\{0,1\}$.
Similarly, the phase factor associated with the $\sigma$-th
encounter averages to ${\rm e}^{-\mu_{\sigma}^2 b t_{\rm
enc}^{\sigma}}$ \cite{Nagao}.  Links and stretches traversed in
opposite directions by $\alpha$ and $\beta$ thus lead to {\it
exponential suppression
  factors} in the contributions of trajectory pairs.

These factors have to be taken into account when evaluating the
average conductance, starting from (\ref{conductance_factored}).
The {\it link integrals} are changed into
\begin{equation}
\int_0^\infty dt_i{\rm e}^{-\frac{N}{T_H}t_i}{\rm e}^{-\mu_i b t_i}=\frac{T_H}{N(1+\mu_i\xi)}\,,
\end{equation}
with $\xi \equiv \frac{T_H}{N}b\propto\frac{B^2}{\hbar}$, whereas for
each {\it encounter} we find an integral
\begin{equation}
\left\langle\int d^{l-1}s d^{l-1}u\frac{1}{\Omega^{l-1}\,t_{\rm enc}^{\sigma}(s,u)}
{\rm e}^{-\frac{N}{T_H}t_{\rm enc}^{\sigma}(s,u)}
{\rm e}^{-\mu_{\sigma}^2 b t_{\rm enc}^{\sigma}(s,u)}
{\rm e}^{{\rm i}\sum_j s_{\sigma j}u_{\sigma j}/\hbar}
\right\rangle=-\frac{N(1+\mu_{\sigma}^2\xi)}{T_H^{l-1}}\,.
\end{equation}
Since the $T_H$'s again mutually cancel, our diagrammatic rules
are changed to give a factor $\frac{1}{N(1+\mu_i\xi)}$ for each
link and a factor $-N(1+\mu_{\sigma}^2\xi)$ for each encounter;
the arising product has to be multiplied with the number of
channel combinations, i.e., $N_1N_2$ for the average conductance.

The same rules carry over to the conductance variance, shot noise,
and correlations in a three-lead geometry. In these cases, $\mu_i$
is equal to 1 if the $i$-th link of the pair ($\alpha$, $\gamma$)
is reverted in ($\beta$, $\delta$), and $\mu_{\sigma}$ counts the
stretches of the $\sigma$-th encounter of ($\alpha$, $\gamma$)
which are  reverted in ($\beta$, $\delta$); the sign of
$\mu_{\sigma}$ is fixed as above.

\subsection{Mean conductance}

For the average conductance, the diagonal contribution,
$\frac{N_1N_2}{N}$, remains unaffected by the magnetic field. The
contribution of Richter/Sieber pairs, $-\frac{N_1N_2}{N^2}$,
obtains an additional factor $\frac{1}{1+\xi}$, since one of the
three links of $\alpha$ in Fig. \ref{fig:RS} or
\ref{fig:conductance_next}a is traversed by $\beta$ in opposite
sense. The next order originates from trajectory pairs as in Fig.
\ref{fig:conductance_next}b-j where arrows indicate the direction
of motion inside the encounters and highlight those links which
are traversed by $\alpha$ and $\beta$ with opposite sense of
motion. The contributions of the families in Fig.
\ref{fig:conductance_next}b, c, d, g, i, j remain unchanged: In
Fig. \ref{fig:conductance_next}b, g no links or encounter
stretches are reverted; for Fig. \ref{fig:conductance_next}c, d,
i, j the number of links with $\mu_i = 1$ and encounters with
$\mu_\sigma^2 = 1$ coincide, meaning that the $\xi$-dependent
factors mutually compensate. The six above families cancel
mutually due to the negative sign for Fig.
\ref{fig:conductance_next}g, i, j. The contributions of Fig.
\ref{fig:conductance_next}e, f, h obtain a factor
$\frac{1}{(1+\xi)^2}$ from two reverted links; due to the negative
sign of Fig. \ref{fig:conductance_next}h, they sum up to
$\frac{N_1N_2}{N^3(1+\xi)^2}$.  We thus find
\begin{eqnarray}
\label{magcond}
\langle\tr(tt^\dagger)\rangle=\frac{N_1N_2}{N}\left\{1-\frac{1}{N(1+\xi)}+\frac{1}{N^2(1+\xi)^2}\right\}
+{\cal
O}\left(\frac{1}{N^2}\right)\,.
\end{eqnarray}
Counting further families of trajectory pairs with the help of a
computer program one is able to proceed to rather high orders in
$\frac{1}{N}$. We then find
\begin{eqnarray}
\label{magcond_num}
\langle\tr(tt^\dagger)\rangle=
\frac{N_1 N_2}{N} \Bigg\{1&-& \frac{1}{N}\frac{1}{(1 + \xi)}
     + \frac{1}{N^2}\frac{1}{(1 + \xi)^2}
-\frac{1}{N^3}\frac{1+2\xi+13\xi^2+4\xi^3+\xi^4}{(1+\xi)^5}
+\frac{1}{N^4}\frac{1+2\xi+49\xi^2+4\xi^3+\xi^4}{(1+\xi)^6}
\nonumber\\
&+&{\cal O}\left(\frac{1}{N^5}\right)\Bigg\}
\end{eqnarray}

As expected, (\ref{magcond}) and (\ref{magcond_num}) interpolate
between the results for the orthogonal case, reached for $B\to 0$
and thus $\xi\to 0$, and the unitary case, formally reached for
$B\to\infty$ and thus $\xi\to\infty$.  The convergence to the
unitary result is non-trivial: The contributions of the families
in Fig.~\ref{fig:conductance_next}c,d,i,j are not affected by a
magnetic field, because all $\xi$-dependent factors cancel.  These
contributions thus survive in the limit $\xi\to\infty$ (i.e., when
the magnetic action becomes much larger than $\hbar$, but the
trajectory deformations due to Lorentz force can still be
disregarded), but vanish in the unitary case (i.e., when the
magnetic field is strong enough to considerably deform the
trajectories).
 The agreement between the limit $\xi\to\infty$ and the unitary
result implies that the contributions of all such families must
sum to zero, for all orders in $\frac{1}{N}$. Order by order,
(\ref{magcond_num}) coincides with the results of
\cite{Weidenmueller}, where the individual coefficients were given
as (rather involved) random-matrix integrals.

\subsection{Conductance variance, shot noise and three-lead
correlations}

For observables determined by families of trajectory quadruplets,
it is convenient to first evaluate the overall contributions of
$d$- and $x$-families {\it per channel combination}. These
contributions, denoted by $D$ and $X$, now depend on the parameter
$\xi$. The contribution of $d$-families reads
\begin{eqnarray}
D&=&\frac{1}{N^2}-\frac{2}{N^3(1+\xi)}
+\frac{1}{N^4}\left[1+\frac{4}{(1+\xi)^2}\right]\nonumber\\
&&+\frac{1}{N^5}\left[-\frac{10}{1+\xi}-\frac{4}{(1+\xi)^3}
-\frac{2(1+4\xi)^2}{(1+\xi)^5}+
\frac{2(1+9\xi)}{(1+\xi)^4}\right]+{\cal
O}\left(\frac{1}{N^6}\right)\,,
\end{eqnarray}
generalizing our previous results (\ref{DX_unit}) and
(\ref{DX_orth}) for the unitary and orthogonal cases. The leading
term, originating from diagonal quadruplets, remains unaffected by
the magnetic field. The second term is due to quadruplets as in
Fig.~\ref{fig:quadruplets}b. Since in these quadruplets, one link
of ($\alpha$, $\gamma$) is time-reversed in ($\beta$, $\delta$),
the corresponding contribution is proportional to
$\frac{1}{1+\xi}$. It is easy to check that the third term
correctly accounts for $d$-quadruplets differing in two
2-encounters, or in one 3-encounter; compare Subsection
\ref{sec:quadruplets} and Fig.~\ref{fig:quadruplets}. The
higher-order terms were again generated by a computer program.

In the overall contribution of {\it $x$-families},
\begin{equation}
X=-\frac{1}{N^3}+\frac{4}{N^4(1+\xi)}+\frac{1}{N^5}
\left[-1-\frac{10}{(1+\xi)^2}-\frac{2(1+4\xi)}{(1+\xi)^4}\right]+{\cal
O}\left(\frac{1}{N^6}\right)\,,
\end{equation}
the term $-\frac{1}{N^3}$ accounts for $x$-quadruplets differing
in a parallel 2-encounter, Fig.~\ref{fig:quadruplets}j.  These
quadruplets are not affected by the magnetic field. All families
responsible for the second term, Fig.~\ref{fig:shot}a-c, display a
Lorentzian field dependence $\frac{1}{1+\xi}$: While
Figs.~\ref{fig:shot}a and \ref{fig:shot}c contain one
time-reversed link and only encounters with $\mu_{\sigma} = 0$,
Fig.~\ref{fig:shot}b involves two time-reversed links and one
encounter with $\mu_{\sigma}^2=1$. The remaining terms were again
found with the help of a computer.

With these values of $D$ and $X$, we obtain, writing out only terms up
to ${\cal O}(N^{-1})$,
\begin{itemize}
\item the {\it conductance variance}
\begin{equation}
\big\langle(\tr(tt^\dagger))^2\big\rangle-\big\langle\tr(tt^\dagger)
\big\rangle^2
=\frac{N_1^2N_2^2}{N^4}\left(1+\frac{1}{(1+\xi)^2}\right)+
-\frac{2N_1^2N_2^2(5+8\xi+4\xi^2)}{N^5(1+\xi)^3}+\frac{2N_1N_2}{N^3(1+\xi)}
+
{\cal O}\left(\frac{1}{N^2}\right)
\,,
\end{equation}
\item the {\it power of shot noise}
\begin{equation}
\big\langle\tr(tt^\dagger)-\tr(tt^\dagger tt^\dagger)\big\rangle
=\frac{N_1^2N_2^2}{N^3}+\frac{N_1N_2(N_1-N_2)^2}{N^4(1+\xi)}
+\frac{N_1^2N_2^2(13+32\xi+16\xi^2+4\xi^3+\xi^4)}{N^5(1+\xi)^4}
-\frac{3N_1N_2}{N^3(1+\xi)^2}
+{\cal O}\left(\frac{1}{N^2}\right)\,,
\end{equation}
\item and {\it current correlations for a cavity with three leads}
\begin{eqnarray}
\langle\tr (t^{(1\to 2)}{t^{(1\to 2)}}^\dagger t^{(1\to
3)}{t^{(1\to 3)}}^\dagger)\rangle
=\frac{N_1N_2N_3(N_2+N_3)}{N^3}+\frac{2N_1N_2N_3(N_1-N_2-N_3)}{N^4(1+\xi)}\nonumber\\
+\frac{N_1N_2N_3\left[(N_2+N_3)(1+\xi)^2(5+2\xi+\xi^2)-2N_1(4+10\xi+3\xi^2)\right]}{N^5(1+\xi)^4}
+{\cal O}\left(\frac{1}{N^2}\right) \,.
\end{eqnarray}
\end{itemize}

At least the higher orders in $\frac{1}{N}$ are new results. In
particular, for the power of shot noise, we do not only obtain the
previously known cancellation of the second term at
$N_1=N_2=\frac{N}{2}$, but also a new field dependence due to the
third term,
\begin{equation}
\big\langle\tr(tt^\dagger)-\tr(tt^\dagger
tt^\dagger)\big\rangle=\frac{N}{16}+\frac{1}{N}\frac{1+8\xi+4\xi^2+4\xi^3+\xi^4}{
16(1+\xi )^4}+{\cal O}\Big( \frac{1}{N^2}\Big) \,;
\end{equation}

\subsection{Ericson fluctuations}

When studying Ericson fluctuations in a weak magnetic field, we
have to deal with {\it two parameters} (apart from the channel
numbers): the scaled energy difference $\epsilon$ {\it and} the
parameter $\xi$ proportional to the squared magnetic field. Our
diagrammatic rules are then changed in a straightforward way. Each
link yields a factor $\frac{1}{N(1+\mu_i\xi-{\rm
i}\eta_{\sigma}\epsilon)}$, whereas each encounter gives
$-N(1+\mu_{\sigma}\xi-{\rm i}\eta_{\sigma}\epsilon)$, to be
multiplied with the number of channel combinations.

As in the orthogonal and unitary cases, the leading contribution
can be attributed to the quadruplets in
Figs.~\ref{fig:covariance}a and \ref{fig:covariance}d; all other
contributions of the same or lower order, including the remaining
families in Fig.~\ref{fig:covariance}, mutually cancel.
Quadruplets as in Fig.~\ref{fig:covariance}a do not feel the
magnetic field, and thus yield
$\frac{N_1^2N_2^2}{N^4(1+\epsilon^2)}$ as shown in Section
\ref{sec:ericson}; here we dropped lower-order corrections due to
the case of coinciding channels.  For the family of quadruplets
depicted in Fig.~\ref{fig:covariance}d, the two links with $\eta_i
= \pm 1$ connecting the two encounters are reversed inside
($\beta$, $\delta$) and thus have $\mu_i=1$. We obtain factors
$\frac{1}{N(1+\xi-{\rm
    i}\epsilon)}$ and $\frac{1}{N(1+\xi+{\rm i}\epsilon)}$ from these
two links, $\frac{1}{N}$ from each of the four remaining link, and
$-N$ from the encounter. Multiplication with the number of channel
combination yields a contribution
$\frac{N_1^2N_2^2}{N^4((1+\xi)^2+\epsilon^2)}$. Ericson
fluctuations in a weak magnetic field are therefore determined as
\begin{equation}
\left\langle G(E)G\left(E+\frac{\epsilon N}{2\pi\overline{\rho}}\right)\right\rangle-\langle G\rangle^2
=\frac{N_1^2N_2^2}{N^4}\left\{\frac{1}{1+\epsilon^2}+\frac{1}{(1+\xi)^2+\epsilon^2}\right\}
+{\cal O}\left(\frac{1}{N}\right)\,.
\end{equation}

\section{Conclusions}

\begin{table}
\begin{center}
\begin{tabular}{|l|ll|c|} \hline
Contribution of each link & simplest case & & $\frac{1}{N}$ \\
\hhline{~---}
& with energy diff. & \!$\propto \epsilon$ and squ. magn. field $\propto \xi$ & $\frac{1}{N(1+\mu_{i}\xi-i\eta_{i}\epsilon)}$ \\
\hline\hline

Contribution of each encounter & simplest case & & $-N$ \\
\hhline{~---}
& with energy diff. & $\propto \epsilon$ and squ. magn. field $\propto \xi$ & $-{N(1+\mu_{\sigma}^2\xi-i\eta_{\sigma}\epsilon)}$ \\
\hline\hline

Total contribution & trajectory pairs & \!\!\vline\; unitary case & $\frac{1}{N}$ \\
\hhline{~~--}
per channel combination & & \!\!\vline\; orthogonal case & $\frac{1}{N+1}$ \\
\hhline{~---}
& $d$-quadruplets & \!\!\vline\; unitary case & $\frac{1}{N^2-1}$ \\
\hhline{~~--}
& & \!\!\vline\; orthogonal case & $\frac{N+2}{N(N+1)(N+3)}$ \\
\hhline{~---}
& $x$-quadruplets & \!\!\vline\; unitary case & $-\frac{1}{N(N^2-1)}$ \\
\hhline{~~--}
& & \!\!\vline\; orthogonal case & $-\frac{1}{N(N+1)(N+3)}$ \\
\hline\hline

Number of channel combinations & trajectory pairs & \!\!\vline\; conductance & $N_1N_2$\\
\hhline{~---}
& $d$-quadruplets & \!\!\vline\; variance of conductance & $N_1^2N_2^2+N_1N_2$\\
\hhline{~~--}
& & \!\!\vline\; shot noise & $N_1N_2N$\\
\hhline{~~--}
& & \!\!\vline\; 3-lead correlations & $N_1N_2N_3$\\
\hhline{~---}
& $x$-quadruplets & \!\!\vline\; variance of conductance & $N_1N_2N$\\
\hhline{~~--}
& & \!\!\vline\; shot noise & $N_1^2N_2^2+N_1N_2$\\
\hhline{~~--}
& & \!\!\vline\; 3-lead correlations & $N_1^2N_2N_3$\\
\hline
\end{tabular}
\end{center}
\caption{Diagrammatic rules determining chaotic quantum transport.
 The table shows link and encounter contributions for all observables
discussed in the present paper. For the simplest cases
(conductance, conductance variance, shot noise, 3-lead correlations),
we have also listed the summed-up contributions of trajectory pairs
and $d$- and $x$-quadruplets and the numbers of channel combinations.}
\label{tab:rules}
\end{table}

A semiclassical approach to transport through chaotic cavities is
established. We calculate mean and variance of the conductance, the
power of shot noise, current fluctuations in cavities with three
leads, and the covariance of the conductance at two different
energies. These observables are dealt with for systems with and
without a magnetic field breaking time-reversal invariance, as well as
in the crossover between these scenarios caused by a weak magnetic
field leading to a magnetic action of the order of $\hbar$.  In contrast
to random-matrix theory, our results apply to {\it individual} chaotic
cavities, and do not require any averaging over ensembles of systems.
Moreover, we go to all orders in the inverse number of channels.

Transport properties are expressed as sums over pairs or
quadruplets of classical trajectories. These sums draw systematic
contributions from pairs and quadruplets whose members differ by
their connections in close encounters, and almost coincide in the
intervening links. The contributions arising from the
topologically different families of quadruplets or pairs are
evaluated using simple and general diagrammatic rules, summarized
in Tab. \ref{tab:rules}. (These rules remain in place
even for observables involving higher powers of the transition matrix,
as shown in Appendix \ref{sec:general}).

Our work shows that, under a set of conditions, 
individual chaotic systems
demonstrate transport properties devoid of any system-specific features
and coinciding with the RMT predictions. An obvious next stage
would be investigation of the system-specific deviations from RMT
observed when these conditions are not met. Previous work
\cite{RS,EssenCond,EssenShot} has already motivated an extension
to the regime where the {\it average dwell time $T_D$ is of
  the order of the Ehrenfest time $T_E$}, i.e.  the duration of the
relevant encounters. Here, the semiclassical approach helped to
settle questions controversial in the random-matrix literature
\cite{EhrenfestRMT}. As shown in \cite{Brouwer,Whitney}, the
leading contributions to the average conductance and the power of
shot noise become proportional to powers of ${\rm e}^{-T_E/T_D}$,
ultimately arising from the exponential decay of the survival
probability. On the other hand, the conductance variance turned
out to be independent of $T_E$ \cite{Brouwer}.

The door is open for a semiclassical treatment of {\it many more
  transport phenomena}, such as quantum decay \cite{Puhlmann}, weak
antilocalization \cite{Zaitsev},
parametric correlations \cite{JapanEssen,KuipersSieber},
the full counting statistics of
two-port cavities, and cavities with more leads.
Extensions to the symplectic symmetry class, along the lines of
\cite{Symplectic,EssenFF}, and to the seven new symmetry classes
\cite{NewSymm} (relevant e.g. for normal-metal/superconductor
heterostructures or quantum chromodynamics) should be within reach. A
generalization to quasi one-dimensional wires would finally lead to a
semiclassical understanding of dynamical localization.

We are indebted to Dmitry Savin and Hans-J{\"u}rgen Sommers (who have
reproduced our prediction (\ref{univshot}) in random-matrix theory
\cite{Dima}); to Piet Brouwer, Phillippe Jacquod,
Saar Rahav, and Robert Whitney for friendly correspondence; to
Taro Nagao, 
Alexander Altland, Ben Simons, Peter Silvestrov, and Martin Zirnbauer for
useful discussions; to Austen Lamacraft for pointing us to
\cite{OneChannel}; and to the Sonderforschungsbereich SFB/TR12 of the
Deutsche Forschungsgemeinschaft and to the EPSRC for financial
support.

\appendix

\section{Trajectory quadruplets vs orbit pairs}\label{sec:orbits}

In this Appendix we will establish a combinatorial 
method for counting families of trajectory quadruplets appearing in
the theory of conductance variance and shot noise.
We will see that trajectory quadruplets  
can be glued together to form orbit pairs, and orbit
pairs can be cut into quadruplets of trajectories. In contrast to
the case of trajectory pairs, see Fig.~\ref{fig:conductance_next},
we shall now need two cuts.

Our approach will be purely topological; e.g., an orbit pair
$({\cal
  A},{\cal B})$ is regarded just as a pair of directed closed lines
with links coinciding in ${\cal A}$ and ${\cal B}$ but differently
connected in the encounters. Similarly, within each quadruplet
$(\alpha,\beta,\gamma,\delta)$ we can assume that the links of
$\alpha,\gamma$ exactly coincide with those of $\beta, \delta$.
Mostly, we can even think of the quadruplets as black boxes with
two left ports $a_1,c_1$ and and two right ports $a_2,c_2$.
Regardless of the actual number of encounters inside, an
$x$-quadruplet can then be treated like a ``dressed'' 2-encounter:
connections $a_1$---$a_2$, $c_1$---$c_2$ in one of the trajectory
pairs are replaced by $a_1$---$c_2$, $c_1$---$a_2$ in the partner
pair, exactly as if a single 2-encounter existed between the
trajectories $\alpha,\gamma$ of the quadruplet. On the other hand,
no change in the connections occurs between the ports in a
$d$-quadruplet, hence it is topologically equivalent to a pair of
dressed links.

We shall consider both the unitary and the orthogonal case. 
In each case, we will use two
slightly different methods to relate trajectory
quadruplets and orbit pairs. 
This will allow us to express
the quantities
$x_m$ and $d_m$ defined in (\ref{dmxm}) through the auxiliary sums
\begin{eqnarray}\label{auxiliaries}
A_m&=&\sum_{\vec{v}}^{M(\vec{v})=m}(-1)^{V(\vec v)}(L(\vec{v})+1){\cal N}(\vec{v})\nonumber\\
B_m&=&\sum_{\vec{v}}^{M(\vec{v})=m}(-1)^{V(\vec v)}{\cal
N}(\vec{v},2)\,,
\end{eqnarray}
where ${\cal N}(\vec{v})$ and ${\cal N}(\vec{v},2)$ are numbers of
structures of orbit pairs (see Subsection \ref{Combi}) and
we have $M(\vv)\equiv L(\vv)-V(\vv)$; these auxiliary sums will be
determined recursively in Subsection \ref{Rec rela} below.

\subsection{Unitary case}

\label{sec:mapping_unitary}

\begin{figure}
\begin{center}
\includegraphics[scale=0.9]{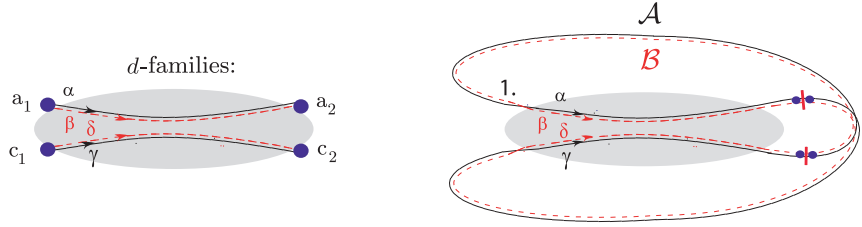}
\end{center}
\caption{Left-hand side: Schematic picture of a $d$-quadruplet of
trajectories
  $\alpha,\gamma$ (full lines), $\beta$, $\delta$ (dashed lines).
  Right-hand side: Outside the ``bubble'', we added connection lines joining
  $(\alpha,\gamma)$ and $(\beta,\delta)$ into periodic orbits ${\cal
  A}$ and ${\cal B}$ with the initial link  indicated by "1.".
  The additional lines of ${\cal A}$ and ${\cal
  B}$ coincide.  }
\label{fig:unitary_glue1}
\end{figure}

\begin{figure}
\begin{center}
  \includegraphics[scale=0.9]{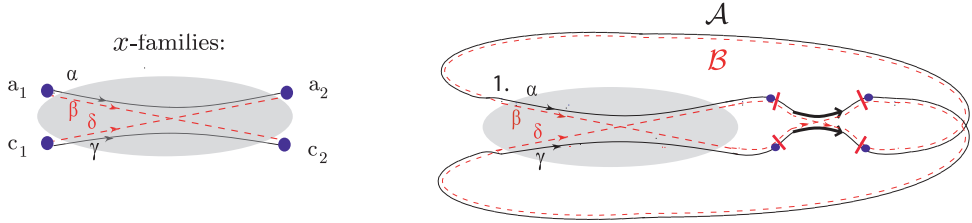}
\end{center}
\caption{Left-hand side: Schematic picture of an $x$-quadruplet of
trajectories
  $\alpha,\gamma$ (full lines), $\beta$, $\delta$ (dashed lines).
  Right-hand side: Outside the ``bubble'', we added connection lines joining
  $(\alpha,\gamma)$ and $(\beta,\delta)$ into periodic orbits ${\cal
    A}$ and ${\cal B}$. The additional lines of ${\cal A}$ and ${\cal
    B}$ differ from each other, and can be viewed as an additional
  2-encounter.}
\label{fig:unitary_glue2}
\end{figure}

To illustrate {\it method I}, let us consider a {\it
$d$-quadruplet} $(\alpha,\beta,\gamma,\delta)$, as on the
left-hand side of Fig.~\ref{fig:unitary_glue1}, and merge $\alpha$
and $\gamma$ into one ``orbit'' ${\cal A}$. We connect the final
point of $\alpha$ to the initial point of $\gamma$ and the final
point of $\gamma$ to the initial point of $\alpha$, as shown on
the right-hand side.  Likewise, $\beta$ and $\delta$ can be glued
together to an ``orbit'' ${\cal
  B}$.\footnote{ As mentioned, $\beta$ and $\delta$ may be
  interchanged if the ingoing and outgoing channels coincide. This has
  no impact on the present considerations.
  The naming of partner trajectories as $\beta$
  and $\delta$  in all figures will be arbitrary.
  } The connection lines added
are the same for $(\alpha,\gamma)$ and for $(\beta,\delta)$: one
connection line joins the coinciding final links of $\alpha$ and
$\beta$ with the coinciding initial links of $\gamma$ and
$\delta$, whereas the second one joins the final links of $\gamma$
and $\delta$ with the initial links of $\alpha$ and $\beta$.  The
orbits ${\cal A}$ and ${\cal B}$ differ in the same encounters as
$(\alpha,\gamma)$ and $(\beta,\delta)$.  To fix one structure for
the orbit pair ${\cal
  A},{\cal B}$, we have to single out one link as the ``first'' and
choose as such the link of ${\cal A}$ created by merging the final
link of $\gamma$ with the initial link of $\alpha$ (indicated by
"1." in Fig.~\ref{fig:unitary_glue1}).

We can revert the above procedure, to obtain families of
$d$-quadruplets from structures of orbit pairs. We first have to
cut both orbits inside the ``initial'' link. This leads to a
trajectory pair with $L(\vec{v})+1$ rather than $L(\vec{v})$
links. We then have $L(\vec{v})+1$ choices for placing a second
cut in any of these links. In each case, we end up with a
trajectory quadruplet. Within this quadruplet, the trajectories
following the first cut through ${\cal
  A}$ and ${\cal B}$ are labelled by $\alpha$ and $\beta$; the
remaining ones are called $\gamma$ and $\delta$. In this way, each
of the ${\cal N}(\vec{v})$ structures of orbit pairs related to a
given $\vec{v}$ gives rise to $L(\vec{v})+1$ families of
$d$-quadruplets with the same $\vec{v}$. The quantities ${\cal
N}_d(\vec{v})$ and $d_m$ characterizing the $d$-families in
(\ref{dmxm}) thus become accessible as
\begin{eqnarray}
{\cal N}_d(\vec{v})&=&(L(\vec{v})+1){\cal N}(\vec{v})\,,\label{dfamnumber}\\
\label{dm_comb_unit}
d_m&=&\sum_{\vec{v}}^{L(\vec{v})-V(\vec{v})=m}(-1)^{V(\vec{v})}(L(\vec{v})+1)
{\cal N}(\vec{v})= A_m\,.
\end{eqnarray}

Let us discuss a few examples. According to (\ref{dm_comb_unit})
the coefficient $d_1$ is determined by orbit pairs with $m=1$
(i.e., only one 2-encounter). Since in the unitary case there are
no such orbit pairs, we have $d_1=0$. The following coefficient
$d_2$ is determined by orbit pairs with $m=2$.  In the unitary
case there are only two such structures, {\it ppi} and {\it pc}
(Fig.~\ref{fig:conductance_next}b and g).  All quadruplets
responsible for the coefficient $d_2$ can be obtained by making
two cuts through these orbit pairs, one through the initial link
which may be chosen arbitrarily.  The second cut can go through
any link; in particular, there are two possibilities for the
second cut in the initial link, before and after the first cut.
That means $L+1$=5 possible positions of the second cut for {\it
ppi}.  These lead to the quadruplets as in
Fig.~\ref{fig:quadruplets}c and d, the reflected version of
Fig.~\ref{fig:quadruplets}d, and quadruplets where either $\alpha$
or $\gamma$ contain two 2-encounters and the other trajectory
contains none. For {\it pc} there are four possible positions for
the second cut, corresponding to Fig.~\ref{fig:quadruplets}e, its
reflected version, and quadruplets where either $\alpha$ or
$\gamma$ contain the full 3-encounter. All quadruplet families
related to a given structure make the same contributions $(-1)^V$
to the coefficient $d_2$, i.e., 1 for those obtained from {\it
ppi} and -1 for those obtained from {\it pc}; we again see that
$d_2=5-4=1$.

To explain {\it method II}, let us now consider {\it
$x$-quadruplets} as on the left-hand side in
Fig.~\ref{fig:unitary_glue2}. On the right-hand side, $\alpha$ and
$\gamma$ are again merged into a periodic orbit ${\cal A}$, and
$\beta$ and $\delta$ are once more merged into ${\cal B}$, by
connection lines leading from the end of one trajectory to the
beginning of the other one. In contrast to the first scenario, the
pair $({\cal A},{\cal B})$ has one further 2-encounter between
these lines, with different connections for the two partner
orbits.  To fix one structure for the latter orbit pair, we take
the initial link of $\alpha$ as the ``first'' link of the orbit
pair. This link is preceded by a ``final'' stretch, which must
belong to the added 2-encounter.  We must therefore reckon with
orbit pairs associated to the vector $\vv^{[\to 2]}$ and whose
final stretches belong to a 2-encounter.  In the notation of
Subsection \ref{sec:conductance_combinatorics}, the number of
structures of such orbit pairs is given by ${\cal N}(\vv^{[\to
2]},2)$.

Each of these structures can be turned back into one family of
$x$-quadruplets, by cutting out the added 2-encounter.
Consequently, there is a one-to-one relation between $x$-families
and the structures of orbit pairs considered. The number of
$x$-families related to $\vv$ and the coefficients $x_m$ defined
in (\ref{dmxm}) are given by
\begin{eqnarray}
{\cal N}_x(\vv)&=&{\cal N}(\vv^{[\to 2]},2)\,,\label{}\\
x_m&=&\sum_{\vv}^{L(\vv)-V(\vv)=m}(-1)^{V(\vv)}{\cal N}(\vv^{[\to
2]},2)\,.
\end{eqnarray}
Rather than over $\vv$, we may sum over $\vv'\equiv\vv^{[\to 2]}$
with $L(\vv')-V(\vv')=(L(\vv)+2)-(V(\vv)+1)=m+1$.  While the
latter sum should be restricted to $\vv'$ with $v_2'>0$, that
restriction may be ignored since $\vv'$ with $v_2'=0$ have ${\cal
N}(\vv',2)=0$.  Using $(-1)^{V(\vv')}=-(-1)^{V(\vv')}$ and
dropping the primes we can express the coefficient $x_m$ as
\begin{equation}
\label{xm_comb_unit}
x_m=-\sum_{\vv}^{L(\vv)-V(\vv)=m+1}(-1)^{V(\vv)}{\cal N}(\vv,2)=
-B_{m+1}\,.
\end{equation}

For instance, the coefficient $x_1$ is determined by orbit pairs
with $L(\vec v)-V(\vec v)=2$. Only one such structure ({\it ppi})
contains 2-encounters and yields a contribution $-1$, whereas {\it
pc} involves only one 3-encounter. We thus find $x_1=-1$, as
already seen previously.

Taken together, Eqs.~(\ref{dm_comb_unit}), (\ref{xm_comb_unit}),
and (\ref{Nvl}) indeed relate the conductance variance to
structures of orbit pairs.

\subsection{Orthogonal case}

\begin{figure}
\begin{center}
  \includegraphics[scale=0.9]{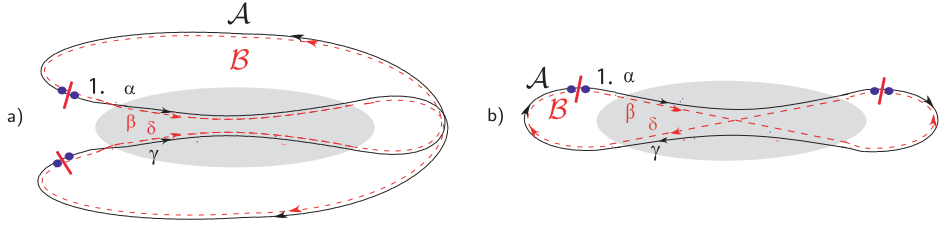}
\end{center}
\caption{Sketches of orbit pairs $({\cal A},{\cal B})$.
  Inside the bubbles: quadruplets of trajectories obtained by cutting
  $({\cal A},{\cal B})$ in the initial link (indicated by "1."), and in one further link.
  In (a) both links are traversed by ${\cal A}$ and ${\cal B}$ with
  the same sense of motion, and the trajectory quadruplet is of type
  $d$. In (b), the second cut is placed in a link traversed with opposite
  senses of motion, and the resulting quadruplet is of type $x$.}
\label{fig:orthogonal_glue1}
\end{figure}

\begin{figure}
\begin{center}
  \includegraphics[scale=0.9]{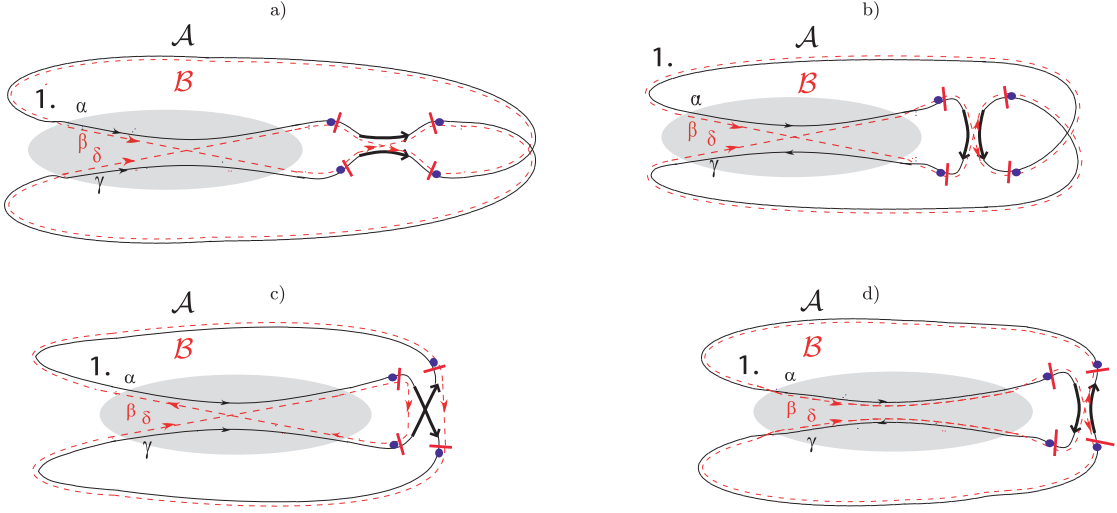}
\end{center}
\caption{Sketches of orbit pairs $({\cal A},{\cal B})$ with one
  2-encounter singled out. This 2-encounter contains the ``final''
  encounter stretch, and is either (a) parallel in both ${\cal A}$ and
  ${\cal B}$, (b) parallel in ${\cal A}$ and antiparallel in ${\cal
    B}$, (c) antiparallel in ${\cal A}$ and parallel in ${\cal B}$, or
  (d) antiparallel in both ${\cal A}$ and ${\cal B}$.}
\label{fig:orthogonal_glue2}
\end{figure}

For time-reversal invariant systems, we must now consider pairs of
orbits ${\cal A}$, ${\cal B}$ differing in encounters whose
stretches are either close or almost mutually time-reversed.  The
sense of traversal of an orbit now being arbitrary we fix the
direction of ${\cal
  B}$ such that ${\cal B}$ traverses the ``initial'' link of ${\cal
  A}$ in the same direction.

We start with {\it method I}, i.e., we {\it cut an orbit pair
$({\cal
    A},{\cal B})$ inside links}, first inside the initial link and
afterwards in an arbitrary link. We have to distinguish two cases,
respectively leading to $d$- and $x$-quadruplets. First assume
that {\it the second cut is placed in a link traversed by ${\cal
A}$ and
  ${\cal B}$ with the same sense of motion}; see
Fig.~\ref{fig:unitary_glue1} or \ref{fig:orthogonal_glue1}a. As in
the unitary case we then obtain a $d$-quadruplet of trajectories;
this quadruplet is highlighted by a grey ``bubble'' in
Fig.~\ref{fig:unitary_glue1} and \ref{fig:orthogonal_glue1}a,
without depicting the encounters. Inside this quadruplet, $\alpha$
is defined as the trajectory following the cut through the initial
link (marked by "1." in
Fig.~\ref{fig:unitary_glue1}-\ref{fig:orthogonal_glue2}).

Now assume that {\it the second cut is placed in a link traversed
by
  ${\cal A}$ and ${\cal B}$ with opposite senses of motion}.  The
resulting quadruplet, shown in the grey bubble in
Fig.~\ref{fig:orthogonal_glue1}b, resembles an $x$-quadruplet,
apart from the directions of motion.  To obtain a true
$x$-quadruplet, one has to revert the directions of motion of two
trajectories in Fig.~\ref{fig:orthogonal_glue1}b, such that all
trajectories point in the same direction as $\alpha$ (i.e., the
trajectory following the cut inside the initial link).

We hence obtain the following relation between orbit pairs and
trajectory quadruplets: By cutting inside links, each of the
${\cal
  N}(\vv)$ structures of orbit pairs related to $\vv$ can be turned
into a family of trajectory quadruplets in $L(\vv)+1$ possible
ways. Through such cuts, all ${\cal N}_d(\vv)$ $d$-families and
all ${\cal
  N}_x(\vv)$ $x$-families are obtained exactly once, since each $d$-
or $x$-family could be inserted inside the bubble in
Fig.~\ref{fig:orthogonal_glue1}a or \ref{fig:orthogonal_glue1}b,
respectively.  We thus have
\begin{equation}
(L(\vec{v})+1){\cal N}(\vec{v})={\cal N}_d(\vec{v})+{\cal
N}_x(\vec{v})\,
\end{equation}
and summation as in the unitary case leads to
\begin{equation}
\label{orthogonal_glue1} A_m=d_m+x_m
\end{equation}
with $A_m$ defined in (\ref{auxiliaries}) and $d_m$, $x_m$ defined
in (\ref{dmxm}).

We now turn to {\it method II} ({\it cutting inside
2-encounters}). We consider orbit pairs ${\cal A},{\cal B}$
containing one more 2-encounter compared with the quadruplets in
question, assuming that the final encounter stretch of the orbits
belongs to the ``added'' 2-encounter; the number of these
structures is  ${\cal N}(\vv^{[\to 2]},2)$. Cutting out the added
2-encounter will create all possible quadruplets associated to
$\vv$, some of them, as we shall see, in several copies.

The added 2-encounter can be either parallel or antiparallel in
the orbit ${\cal A}$ as well as in its partner ${\cal B}$. This
leads to four different possibilities, depicted by arrows on white
background in Fig.~\ref{fig:orthogonal_glue2}a-d.  First suppose
that the encounter in question is antiparallel in both ${\cal A}$
and ${\cal B}$, as in Fig.~\ref{fig:orthogonal_glue2}d.  This is
possible only if the ports of this encounter are connected in the
same way in ${\cal A}$ and ${\cal B}$, up to time reversal; one
can easily check that all other connections would lead to either
${\cal A}$ or ${\cal
  B}$ decomposing into several disjoint orbits.  The connections
outside the encounter, as depicted in the bubble in
Fig.~\ref{fig:orthogonal_glue2}d, thus resemble $d$-quadruplets
and can be turned into true $d$-quadruplets if we revert the sense
of motion on some trajectories.\footnote{ Rather than reverting
directions of
  motion, we could also identify the initial points of all
  trajectories inside the bubble with ingoing leads, and the final
  points with outgoing leads, loosing the identification of the two
  sides of our bubble with the two openings of the cavity. This would
  entail a different mapping between orbit pairs and trajectory
  quadruplets, but not affect the following results.}  Any $d$-family
could be substituted for the bubble in Fig.
\ref{fig:orthogonal_glue2}d.  Thus, cutting through 2-encounters
of the kind in Fig. \ref{fig:orthogonal_glue2}d produces all
possible families of $d$-quadruplets.

In the three other cases, Fig.~\ref{fig:orthogonal_glue2}a-c, the
remaining connections have to be of type $x$, up to the sense of
motion on some trajectories, since connections of type $d$ would
lead to decomposing orbits.  To better understand these cases, it
is helpful to view the corresponding $x$-quadruplets as a
``dressed'' 2-encounters. Then, the orbit pairs of
Fig.~\ref{fig:orthogonal_glue2}a-c are topologically equivalent to
the simple diagrams in Fig.~\ref{fig:conductance_next}. The orbit
pair in Fig.~\ref{fig:orthogonal_glue2}a is of the type {\it ppi}
whereas the pairs in Fig.~\ref{fig:orthogonal_glue2}b and c are
both of the type {\it api}. (In Fig. \ref{fig:orthogonal_glue2}b
the initially parallel encounter is identified with the added
encounter, and the initially antiparallel encounter is identified
with the $x$-quadruplet, whereas the situation is opposite in
Fig.~\ref{fig:orthogonal_glue2}c.) Any family of type $x$ could be
substituted for the bubbles in each of the
Figs.~\ref{fig:orthogonal_glue2}a-c, and can therefore be obtained
by cutting three different structures of orbit pairs.

We have thus seen that by cutting through 2-encounters of the
${\cal
  N}(\vv^{[\to 2]},2)$ structures of orbit pairs considered, we obtain
each family of $d$-quadruplet once and once only, whereas each
$x$-family is produced by three different structures.  We
therefore have
\begin{equation}
{\cal N}(\vv^{[\to 2]},2)={\cal N}_d(\vv)+3{\cal N}_x(\vv)\,,
\end{equation}
and, by summing over $\vv$ as in Subsection
\ref{sec:mapping_unitary},
\begin{equation}
\label{orthogonal_glue2} -B_{m+1}=d_m+3x_m\,.
\end{equation}
 Eqs. (\ref{orthogonal_glue1}) and (\ref{orthogonal_glue2}) form
a system of equations for the coefficients $d_m$, $x_m$, with the
solution
\begin{subequations}
\begin{eqnarray}
\label{dm_comb_orth}
d_m&=&\frac{3}{2}A_m+\frac{1}{2}B_{m+1}\\
\label{xm_comb_orth} x_m&=&-\frac{1}{2}A_m-\frac{1}{2}B_{m+1}
\end{eqnarray}
\end{subequations}

\subsection{Recursion relations}\label{Rec rela}

We must calculate the auxiliary sums $A_m,B_m$ defined in
(\ref{auxiliaries}) for both the unitary and the orthogonal cases.
We start from the {\it recursion for the number of
  structures ${\cal N}(\vec{v})$} already used to evaluate the average
conductance in Subsection \ref{sec:conductance_combinatorics}
\begin{equation}
\label{recursion2_again} {\cal N}(\vec{v},2)-\sum_{k\geq 2}{\cal
N}(\vec{v}^{[k,2\to k+1]},k+1)
=\left(\frac{2}{\beta}-1\right){\cal N}(\vec{v}^{[2\to]})\,;
\end{equation}
see Eq.~(\ref{recursion2}).  This time, (\ref{recursion2_again})
has to be multiplied not with $(-1)^{V(\vv)}$, but with
$(-1)^{V(\vec{v})}L(\vec{v})=-(-1)^{V(\vec{v}^{[k,2\to
    k+1]})}(L(\vec{v}^{[k,2\to
  k+1]})+1)=-(-1)^{V(\vec{v}^{[2\to]})}(L(\vec{v}^{[2\to]})+2)$.  If
we subsequently sum over all $\vec{v}$ with $M(\vec{v})=m$, our
recursion turns into
\begin{eqnarray}
&&\sum_{\vv}^{M(\vv)=m}\left\{(-1)^{V(\vv)}L(\vv){\cal N}(\vv,2)+\sum_{k\geq 2}(-1)^{V(\vv^{[k,2\to k+1]})}(L(\vv^{[k,2\to k+1]})+1){\cal N}(\vv^{[k,2\to k+1]},k+1)\right\}\nonumber\\
&&=-\left(\frac{2}{\beta}-1\right)\sum_{\vv}^{M(\vec{v})=m}(-1)^{V(\vv^{[2\to]})}(L(\vv^{[2\to]})+2){\cal
N}(\vv)\,.
\end{eqnarray}
Changing the summation variables as in Subsection
\ref{sec:conductance_combinatorics}, we find
\begin{eqnarray}
&&\sum_{\vv}^{M(\vv)=m}\left\{(-1)^{V(\vv)}L(\vv){\cal N}(\vv,2)+\sum_{k\geq 2}(-1)^{V(\vv)}(L(\vv)+1){\cal N}(\vv,k+1)\right\}\nonumber\\
&&=-\left(\frac{2}{\beta}-1\right)\sum_{\vv}^{M(\vec{v})=m-1}(-1)^{V(\vv)}(L(\vv)+2){\cal
N}(\vv)\,.
\end{eqnarray}
Using ${\cal N}(\vv,2)+\sum_{k\geq 2}{\cal N}(\vv,k+1)={\cal
N}(\vv)$, we can simplify the left-hand side to
$\sum_{\vv}^{M(\vv)=m}(-1)^{V(\vv)}(L(\vv)+1){\cal
  N}(\vv)-\sum_{\vv}^{M(\vv)=m}(-1)^{V(\vv)}{\cal N}(\vv,2)=A_m-B_m$,
whereas the right-hand side turns into
$-\left(\frac{2}{\beta}-1\right)(A_{m-1}+c_{m-1})$ with
$c_{m-1}=\sum_{\vv}^{M(\vv)=m-1}(-1)^{V(\vv)}{\cal N}(\vv)$, see
Eq. (\ref{cm}).  We thus find a first relation between $A_m$ and
$B_m$,
\begin{equation}
\label{relation1}
A_m-B_m=-\left(\frac{2}{\beta}-1\right)(A_{m-1}+c_{m-1})\,.
\end{equation}

A {\it second relation between $A_m$ and $B_m$} follows from the
recursion \cite{EssenFF} \footnote{ Eq. (\ref{recursion3}) follows
  from the special case $l=3$ of Eqs. (42), (54) in \cite{EssenFF}. To
  understand the equivalence to \cite{EssenFF}, we need the identity
  ${\cal N}(\vv^{[3\to 1,1]},1)\stackrel{(*)}{=}{\cal N}(\vv^{[3\to
    1]})\stackrel{\mbox{(\ref{Nvl})}}{=} \frac{L(\vv^{[3\to
      1]})}{v_1^{[3\to 1]}}{\cal N}(\vv^{[3\to 1]},1)\stackrel{(*)}{=}
  \frac{L(\vv^{[3\to 1]})}{v_1^{[3\to 1]}}{\cal N}(\vv^{[3\to]})
  \stackrel{(**)}{=}(L(\vv^{[3\to]})+1){\cal N}(\vv^{[3\to]})$,
  following from $(*)$ ${\cal N}(\vv',1)={\cal N}(\vv'^{[1\to ]})$
  (see the footnote in Subsection
  \ref{sec:conductance_combinatorics}), Eq. (\ref{Nvl}) of the present
  paper, and $(**)$ the fact that $v_1=0$ and thus $v^{[3\to
    1]}_1=1$.}
\begin{equation}
\label{recursion3} {\cal N}(\vv,3)-\sum_{k\geq 2}{\cal
N}(\vv^{[k,3\to k+2]},k+2)=2\left(\frac{2}{\beta}-1\right){\cal
N}(\vv^{[3\to2]},2) +\frac{2}{\beta}(L(\vv^{[3\to]})+1){\cal
N}(\vv^{[3\to]})\,,
\end{equation}
which has to be multiplied with
$(-1)^{V(\vv)}=-(-1)^{V(\vv^{[k,3\to
    k+2]})}=(-1)^{V(\vv^{[3\to 2]})}=-(-1)^{V(\vv^{[3\to]})}$ and
summed over all $\vv$ with $M(\vv)=m$.  It is easy to see that the
changed vectors $\vv^{[k,2\to k+1]}, \vv^{[3\to 2]}, \vv^{[3\to]}$
in the arguments of ${\cal N}$ have $M(\vv^{[k,3\to k+2]})=m$,
$M(\vv^{[3\to 2]})=m-1$, and $M(\vv^{[3\to]})=m-2$. Again
transforming the sums over $\vv$ into sums over the arguments of
${\cal N}$, we are led to
\begin{eqnarray}
\label{recursion3_summed}
&&\sum_{\vv}^{M(\vv)=m}(-1)^{V(\vv)}\left\{{\cal N}(\vv,3)+\sum_{k\geq 2}{\cal N}(\vv,k+2)\right\}\nonumber\\
&&=2\left(\frac{2}{\beta}-1\right)\sum_{\vv}^{M(\vv)=m-1}(-1)^{V(\vv)}{\cal
N}(\vv,2)
-\frac{2}{\beta}\sum_{\vv}^{M(\vv)=m-2}(-1)^{V(\vv)}(L(\vv)+1){\cal
N}(\vv)\,.
\end{eqnarray}
Eq. (\ref{recursion3_summed}) can be simplified if we use ${\cal
  N}(\vv,3)+\sum_{k\geq 2}{\cal N}(\vv,k+2)={\cal N}(\vv)-{\cal
  N}(\vv,2)$, and recall the definitions of $A_m$, $B_m$ and $c_m$, to
get
\begin{equation}
\label{relation2}
c_m-B_m=2\left(\frac{2}{\beta}-1\right)B_{m-1}-\frac{2}{\beta}A_{m-2}
\end{equation}

\subsection{Results}
In the {\it unitary case} the two relations between $A_m$ and
$B_m$, Eqs. (\ref{relation1}) and (\ref{relation2}) yield a
recursion for both quantities.  In the {\it
  unitary case}, $\frac{2}{\beta}-1=0$, $c_m=0$, the two equations
simplify to
\begin{eqnarray}
A_m-B_m=0,\qquad -B_m=-A_{m-2}\,,
\end{eqnarray}
i.e., $A_m$ and $B_m$ coincide and depend only on whether $m$ is
even or odd.
Since in the unitary case
$d_m=A_m,\,x_m=-B_{m+1}$, see (\ref{dm_comb_unit}), and since the
initial values $d_1=0,d_2=1$ are already established,  we come to
the expressions (\ref{xd unitary}) for $x_m,d_m$.

In the {\it orthogonal case} $\frac{2}{\beta}-1=1$, $c_m=(-1)^m$,
Eqs.~(\ref{relation1}) and (\ref{relation2}) take the form
\begin{subequations}
\begin{eqnarray}
\label{orth1}
A_m-B_m&=&-A_{m-1}+(-1)^m\\
\label{orth2} (-1)^m-B_m&=&2B_{m-1}-2A_{m-2}\,.
\end{eqnarray}
\end{subequations}
Eliminating $B_m$ and $B_{m-1}$ in (\ref{orth2}) with the help of
(\ref{orth1}), we find a recursion for $A_m$,
\begin{equation}
A_m=-3A_{m-1}\,.
\end{equation}
An initial condition is provided by $A_1=-3$, which accounts for
the Richter/Sieber family of trajectory pairs with one
2-encounter, $V=1$, $L=2$ and $m=1$. We thus obtain
\begin{equation}
A_m=(-3)^m,\qquad
B_m\stackrel{\mbox{(\ref{orth1})}}{=}A_m+A_{m-1}+(-1)^{m-1}=(-1)^m(2\cdot
3^{m-1}-1)\,.
\end{equation}
After that, using (\ref{dm_comb_orth}), (\ref{xm_comb_orth}) we
arrive at the result (\ref{xd orthogonal}) of the main text.

\section{An algorithm for counting trajectory pairs and
quadruplets}

\label{sec:numerics}

In our approach transport properties phenomena are related to
trajectory pairs or quadruplets differing in encounters; we showed that all
topologically different families of pairs or quadruplets
can be found using the corresponding structures of orbit pairs
considered in the theory of spectral fluctuations
\cite{EssenFF}. The number of structures to be considered grows
exponentially with the order of approximation, and in
absence of an analytic formula (in particular, for Ericson
fluctuations or crossover in a magnetic field)
can be evaluated only
with the help of a computer. Here we describe an algorithm for the
systematic generation of orbit pairs.

Assume time-reversal invariant dynamics, and let $\mathcal A$ be a periodic orbit and
$\mathcal B$ its partner obtained by reconnections in a set of encounters associated to the vector
$\vec v=\left( v_{2},v_{3},v_{4},\ldots \right) $. Let us then number all the
$L=L(\vec{v})$ encounter stretches in $\mathcal{A}$ in order of
traversal. Each encounter has two sides which can be arbitrarily
named left and right such that the two ends of the $i$ -th encounter
stretch can be called its left ($i_{l}$) and right ($i_{r}$) ports.
The direction of traversal of the encounter stretches in $\mathcal{A}$
will be denoted by another vector $\overrightarrow{\sigma }=\left( \sigma _{1}\sigma
  _{2}\ldots \sigma _{L}\right)$.
Here $\sigma_i$ is equal to $1$ if the $i$
-th stretch is traversed from the left to the right such that $i_{l}$
and $i_{r}$ are its entrance and exit ports respectively, and equal to -1 in
the opposite case.
There can be $2^{L}$ different $ \vec{\sigma }$. However, only relative
direction of motion within encounters is physically meaningful: we
can, e.g., assume that the first stretch of each of the $V\left(
  \vec{v}\right) $ encounters is passed from the left to the right.
This leaves $2^{L-V}$ physically different $\vec{\sigma }$.
 Each orbit link connects the exit port of an
encounter stretch of $ \mathcal{A}$ with the entrance port of the
following stretch. Both of these ports can be left or right which
makes four combinations possible, $i_{r}\to (i+1)_{l},\quad
i_{r}\to (i+1)_{r},\quad i_{l}\to (i+1)_{l},\quad
i_{l}\to(i+1)_{l}$, the sense of the link traversal in $\mathcal
A$ being in all cases $i\to i+1$. These choices are uniquely fixed
by the vector $\vec{\sigma }$.

In the partner orbit $\mathcal{B}$ the left port $i_{l}$ will be
connected by an encounter stretch not with the right port $i_{r}$
but with some other right port $ f(i)_{r}$. The set of
reconnections of all ports can be written as a permutation
\begin{equation}
P_{\rm enc}={{1,\quad 2,\ldots L}\choose{f(1),f(2)\ldots f(L)}}
\end{equation}
 in which
the upper and lower lines refer to the left and right encounter
ports, correspondingly.  Since reconnections are possible only
within encounters, $P_{\rm enc}$ must consist of as many
independent cyclic permutations (cycles) as there are encounters
in the orbit $\cal A$, with each cycle of $l$ elements
corresponding to an $l$-encounter. The links of the orbit
$\mathcal{B}$ are unchanged compared to $\mathcal{A}$, but may be
passed with the opposite sense. The orbit $\mathcal{B}$ may exist
in two time reversed versions; we shall choose the one in which
the encounter stretch with the port $1_{l}$ at its left is passed
from the left to the right. This choice made, the sequence of
visits of all ports and the direction of traversal of all
encounter stretches and links in $ \mathcal{B}$ become uniquely
fixed by the permutation $P_{\rm enc}$ and the the vector
$\vec{\sigma}$. Indeed, let us start from $1_{l}$, move along the
encounter stretch arriving at the right \ port $f(1)_{r}$ and then
traverse the link attached.  Where we move next in $\mathcal{B}$
depends on the ports connected by this link in the original orbit
$\mathcal{A}$:
\begin{enumerate}
\item  $f(1)_{r} \to\ \left[ f(1)+1\right] _{l}$ . The sense of
traversal of the link in $\mathcal{B}$ is the same as in
$\mathcal{A}$. The next encounter stretch is traversed from the
left to the right leading  from the port $p_{2}^{\prime }=\left[
f(1)+1\right] _{l}$  to $\ \left[ f(p_{2}^{\prime })\right] _{r}$.

\item $f(1)_{r}\to \left[ f(1)+1\right] _{r}.$ The sense of
traversal of the link in $\mathcal{B}$ is the same as in
$\mathcal{A}$. The next encounter stretch is traversed from the
right to the left leading from $ p_{2}^{\prime }=\left[
f(1)+1\right] _{r}$ to $\left[ f^{-1}(p_{2}^{\prime }) \right]
_{l}$, i.e., to the element of the upper row in $P_{\rm enc}$
corresponding to $ f(1)+1$ in the lower row.

\item $\left[ f(1)-1\right] _{l}\to f(1)_{r}\ .$ The link is
traversed in $\mathcal{B}$ in the direction opposite to
$\mathcal{A}$ leading from $f(1)_{r}$ to $p_{2}^{\prime }=\left[
f(1)-1\right] _{l}$. The next encounter stretch leads from the
left port $p_{2}^{\prime }$ to the right port $\left[
f(p_{2}^{\prime })\right] _{r}.$

\item $\left[ f(1)-1\right] _{r}\to f(1)_{r}.$ The link is
traversed in $\mathcal{B}$ in the direction opposite to
$\mathcal{A}$ leading from $\ f(1)_{r}$ to $p_{2}^{\prime }=\left[
f(1)-1\right] _{r}$.  The next encounter stretch leads from the
right port $p_{2}^{\prime }$ to the left port $\left[
f^{-1}(p_{2}^{\prime })\right] _{l}$.
\end{enumerate}

Continuing our way we eventually return to the starting port
$1_l$. If that return occurs before all $2L$ encounter ports are
visited the combination $P_{\rm enc},\vec \sigma$ has to be
discarded since it leads to a partner consisting of several
disjoint orbits, a so called pseudo-orbit.  Otherwise we have
found a structure of the periodic orbit pair $\mathcal{(A,B)}$
with the encounter set $\vec v$ and established the port sequence
in $\mathcal{B}$ as well as the sense of traversal of all its
links and encounter stretches. Running through all $
L!/\prod_{l\geq 2}l^{v_{l}}v_{l}!$ permutations associated to
$\vec{v}$ and through all $2^{L-V}$ essentially different $
\vec{\sigma }$ we find all $N(\vec{v})$ structures of the orbit
pairs.

Suitable cuts yield all trajectory doublets and quadruplets
relevant for quantum transport. E.g., if we cut $\mathcal{A}$ and
$\mathcal{B}$ in the initial link (i.e. in the link preceding the
port $1_l$) we obtain a trajectory pair contributing to the
conductance; the numbers $\mu_i,\mu_\sigma$ needed for calculating
conductance in a magnetic field are obtained by counting the
number of links and encounter stretches changing their direction
in $\mathcal B$ compared with $\mathcal A$.

The trajectory quadruplets contributing to the conductance
covariance and other properties are obtained, in accordance with
our method I (see Appendix \ref{sec:orbits}), by cutting the pair $\mathcal{(A,B)}$ twice, in the
initial link and in any of the $L$ orbit links producing thus
$L+1$ quadruplets per orbit pair. If the second cut goes through a
link preserving its sense of traversal the result is a
$d$-quadruplet, otherwise it is an $x$-quadruplet. (We remind the
reader that in the last case the sense of traversal of the
trajectories $\gamma,\delta$ is {\it   changed to the opposite}
compared with  the periodic orbits; this has to be taken into
account in calculation of the transport properties in the magnetic
field.) Method II of producing quadruplets consists of cutting a
2-encounter out of the orbit pair $\mathcal{(A,B)}$ which is
 possible only if the encounter set $\vec v$ contains at
least one 2-encounter, i.e. $v_2>0$; the resulting quadruplet will
be characterized by the encounter set $\vec v'=\vec v^{[2\to]}$.

The relatively trivial case when time reversal is absent can be
treated by choosing $\sigma _{i}=+1,\,\,i=1,\ldots ,L$ (all
encounter stretches are traversed from the left to the right, and
all links are attached to the ports like $i_{r}\to \left(
i+1\right) _{l}).$

\section{Diagrammatic rules for arbitrary multiplets of trajectories}

\label{sec:general}

Our semiclassical techniques can be expanded to a huge class of
transport problems, for cavities with arbitrary numbers of leads,
and observables involving arbitrary powers of transition matrices.
We then have to evaluate (sums of) general products of the type
\begin{equation}
\label{general_observable} \langle Z\rangle=
\left\langle\prod_{m=1}^Mt_{a_1^{(m)}a_2^{(m)}}t_{a_1^{(m)}a_2^{(\Pi(m))}}^*\right\rangle\,,
\end{equation}
with $a_1^{(1)},\ldots,a_1^{(M)},a_2^{(1)},\ldots,a_2^{(M)}$
denoting $2M$ mutually different channel indices associated to any
of the attached leads, and $\Pi$ a permutation of $1,2,\ldots,M$.
Using the semiclassical transition
amplitudes (\ref{transition}), we express $\langle Z\rangle$ as a
sum over multiplets of trajectories $\alpha_k,\beta_k$ connecting
channels as $\alpha_m\big(a_1^{(m)}\to a_2^{(m)}\big)$,
$\beta_m\big(a_1^{(m)}\to a_2^{(\Pi(m))}\big)$,
\begin {equation}
\label{multiple_sum} \langle
Z\rangle=\frac{1}{T_H^M}\left\langle\sum_{\alpha_1,\ldots,\alpha_M\atop
    \beta_1,\ldots,\beta_M} A_{\alpha_1}\ldots
  A_{\alpha_M}A_{\beta_1}^*\ldots A_{\beta_M}^* {\rm e}^{{\rm
      i}(S_{\alpha_1}+\ldots
    +S_{\alpha_M}-S_{\beta_1}-\ldots-S_{\beta_M})/\hbar}\right\rangle\,.
\end {equation}
Contributions to (\ref{multiple_sum}) arise from multiplets of
trajectories where $\beta_1,\ldots,\beta_M$ either coincide with
$\alpha_1,\ldots,\alpha_M$, or differ from the latter trajectories
only inside close encounters in phase space. $\langle Z\rangle$
thus turns into a sum over families of multiplets characterized by
a vector $\vec{v}$.

Proceeding as in Subsections \ref{sec:conductance_general} and
\ref{sec:quadruplets}, we represent the contribution of each
family as a sum over the trajectories $\alpha_1,\ldots,\alpha_M$
and an integral over the density of stable and unstable
coordinates,
\begin{equation}
\langle Z\rangle \big|_{\rm
fam}=\frac{1}{T_H^M}\left\langle\sum_{\alpha_1,\ldots,\alpha_M}|A_{\alpha_1}|^2\ldots|A_{\alpha_M}|^2
\int d^{L-V}s\,d^{L-V}u\, w(s,u) {\rm
e}^{-\frac{N}{T_H}\left(\sum_{i=1}^{L+M}t_i+\sum_{\sigma=1}^V
t_{\rm enc}^\sigma(s,u)\right)} {\rm e}^{{\rm
i}\sum_{\sigma,j}s_{\sigma j}u_{\sigma j}/\hbar}\right\rangle\,;
\end{equation}
here, $w (s, u)$ is obtained by integrating
$\{\Omega^{L-V}\prod_{\sigma=1}^Vt_{\rm enc}^\sigma(s,u)\}^{-1}$
over the durations of all links except the final link of each of
the $M$ trajectories. Doing the sum over $\alpha_k$ with the
Richter/Sieber rule, we find the same link and encounter integrals
as before, and thus a contributions $\frac{1}{N}$ from each link
and a contribution $-N$ from each encounter.

The same rule applies for products slightly different from
(\ref{general_observable}). First, if some of the channels
$a_1^{(1)},\ldots,a_1^{(M)}$ or some of the channels
$a_2^{(1)},\ldots,a_2^{(M)}$ coincide (as in many examples studied
in the main part) some of the subscripts in
(\ref{general_observable}) will appear not twice but $4, 6, 8,
\ldots$ times. We then have to consider all possible ways to pair
these subscripts. In the spirit of Wick's theorem, each of these
possibilities contributes separately. (Note that, if a subscript
appears an odd number of times, the corresponding product cannot
be related to multiplets of trajectories, and may be expected to
vanish after averaging over the energy). Second, in the orthogonal
case the two subscripts of one $t$ or $t^*$ in
(\ref{general_observable}) may be interchanged without affecting
the final result; the corresponding trajectory is then reverted in
time.

With these rules, one can evaluate a huge class of observables
relevant for quantum transport. For each single application, only
the counting of families remains to be mastered.

\section{Spectral statistics revisited}

\label{sec:spectral}

We here want to reformulate our previous results
on spectral statistics \cite{EssenFF} in the present language of
diagrammatic rules. In contrast to \cite{EssenFF}, we start from the
level staircase $N(E)$, defined as the number of energy
eigenvalues below $E$. $N(E)$ can be split into a smooth local average
$\overline{N}(E)$ and an fluctuating part $N_{\rm osc}(E)$ describing
fluctuations around that average. We want to study the two-point
correlation function of $N_{\rm osc}(E)$
\begin{equation}\label{staircorr}
C(\epsilon)=\left\langle N_{\rm
osc}\left(E+\frac{\epsilon}{2\pi\overline{\rho}}\right) N_{\rm
osc}\left(E-\frac{\epsilon}{2\pi\overline{\rho}}\right)\right\rangle\,.
\end{equation}
The latter correlator yields the spectral form factor $K(\tau) =
\frac{1}{\pi}\int_{-\infty}^{\infty}d\epsilon \,{\rm e}^{2{\rm i}\epsilon\tau}R(\epsilon)$ through
the identity $R(\epsilon)=\frac{1}{\overline{\rho}^2}\left\langle \frac{d N_{\rm
      osc}}{d E}\left(E+\frac{\epsilon}{2\pi\overline{\rho}}\right) \frac{d
    N_{\rm osc}}{dE}\left(E-\frac{\epsilon}{2\pi\overline{\rho}}\right)\right\rangle
=-\pi^2\frac{d^2 C}{d\epsilon^2}$. To check on the last member of
the foregoing chain of equations one must write the average
$\langle \cdot \rangle$ in (\ref{staircorr}) as an integral over
the center energy $E$, take its second derivative by $\epsilon$
and integrate by parts the terms containing $N''N$ in the
integrand.

In the semiclassical limit, Gutzwiller's trace formula determines
$N_{\rm osc}(E)$ as a sum over periodic orbits ${\cal A}$ of
arbitrary period $T_{\cal A}(E)$, $N_{\rm
osc}(E)=\frac{1}{\pi}{\rm
  Im}\sum_{\cal A}F_{\cal A}{\rm e}^{{\rm i}S_{\cal A}(E)}$; here,
$F_{\cal A}$ depends on the stability matrix $M_{\cal A}$ and the
Maslov index $\mu_{\cal A}$ of ${\cal A}$ as $F_{\cal
  A}=\frac{1}{\sqrt{|\det(M_{\cal A}-1)|}}{\rm e}^{{\rm i}\mu_{\cal
    A}\frac{\pi}{2}}$, and $S_{\cal A}(E)$ is the classical action of
${\cal A}$ at energy $E$. The correlation function $C(\epsilon)$
turns into a double sum over periodic orbits ${\cal A}$ and ${\cal
B}$,
\begin{equation}
\label{doublesum_C} C(\epsilon)=\frac{1}{2\pi^2}{\rm
Re}\left\langle\sum_{{\cal A}, {\cal B}}F_{\cal A} F_{\cal
B}^*{\rm e}^{{\rm i}(S_{\cal A}(E)-S_{\cal B}(E))/\hbar}{\rm
e}^{{\rm i}\frac{T_{\cal A}(E)+T_{\cal
B}(E)}{T_H}\epsilon}\right\rangle\,,
\end{equation}
where we have used $S_{\cal
A}(E+\frac{\epsilon}{2\pi\overline{\rho}}) \approx S_{\cal
A}(E)+T_{\cal
  A}(E)\frac{\epsilon}{2\pi\overline{\rho}}$, $S_{\cal
  B}(E-\frac{\epsilon}{2\pi\overline{\rho}}) \approx S_{\cal
  B}(E)-T_{\cal B}(E)\frac{\epsilon}{2\pi\overline{\rho}}$ and
$T_H=2\pi\hbar\overline{\rho}$.

To evaluate the contribution to (\ref{doublesum_C}) resulting from
a given structure of orbit pairs differing in encounters (see
Subsection \ref{sec:conductance_general}), we replace the sum over
${\cal B}$ by an integral over a density $w(s, u)$ of phase-space
separations inside ${\cal A}$.  Similarly as for transport
$w(s,u)$ is defined as the integral of
$\frac{1}{\Omega^{L-V}\prod_{\sigma=1}^Vt_{\rm
    enc}^\sigma(s,u)}$ over all piercing times; integration over the
first piercing time leads to multiplication with the orbit period,
whereas the remaining integrals can be transformed into integrals
over all link durations but one.  Approximating $F_{\cal B}\approx
F_{\cal
  A}$ and $T_{\cal B}\approx T_{\cal A}$, we find
\begin {equation}
\label{integral_C} C(\epsilon)\big|_{\rm fam}=\frac{1}{2\pi^2
L}\left\langle\sum_{\cal
    A}|F_{\cal A}|^2 \int d^{L-V}s\int d^{L-V}u\,w(s,u){\rm e}^{{\rm
      i}(S_{\cal A}(E)-S_{\cal B}(E))/\hbar} {\rm e}^{2{\rm i}T_{\cal
      A}\epsilon/T_H}\right\rangle\,.
\end {equation}
We here divided out $L$, because for each orbit pair any of the
$L$ links may be chosen as the ``first''; without this division,
each pair would be counted $L$ times. The sum over ${\cal A}$ can
now be done using the sum rule of Hannay and Ozorio de Almeida
\cite{HOdA},
\begin{eqnarray}
\sum_{\cal A}|F_{\cal A}|^2\delta(T-T_{\cal A})=\frac{1}{T}
\;\;\Rightarrow\;\;\sum_{\cal A}|F_{\cal A}|^2\,(\cdot) = \int
dT\frac{1}{T}\,(\cdot)\,.
\end{eqnarray}
The multiplication with the orbit period is thus replaced by
integration over the period or, equivalently, over the duration of
the remaining link. Writing $T_{\cal A}=\sum_{i=1}^L
t_i+\sum_{\sigma
  =1}^Vl_\sigma t_{\rm enc}^\sigma(s,u)$, we can now split
(\ref{integral_C}) into the prefactor $\frac{1}{2\pi^2 L}$, an
integral
\begin{equation}
\int_0^\infty d t_i {\rm e}^{2{\rm
i}t_i\epsilon/T_H}=-\frac{T_H}{2{\rm i \epsilon}}
\end{equation}
for each link and an integral
\begin{equation}
\left\langle\int d^{l-1}s d^{l-1}u\frac{1}{\Omega^{l-1}\,t_{\rm
enc}^{\sigma}(s,u)} {\rm e}^{2{\rm i}l_\sigma t_{\rm
enc}^{\sigma}(s,u)\epsilon/T_H} {\rm e}^{{\rm i}\sum_j s_{\sigma
j}u_{\sigma j}/\hbar}\right\rangle = \frac{2l_\sigma{\rm
i}\epsilon}{T_H^l}
\end{equation}
for each encounter; to ensure convergence we assume $\epsilon$ to
have an infinitesimal positive imaginary part. Since all $T_H$'s
cancel, we obtain a factor $-\frac{1}{2{\rm i}\epsilon}$ for each
link, and a factor $2l{\rm i}\epsilon$ for each $l$-encounter.
The overall product reads
\begin{equation}
C(\epsilon)\big|_{\rm fam}=\frac{1}{2\pi^2 L}(-1)^L\prod_l
l^{v_l}{\rm Re}(2{\rm i}\epsilon)^{V-L}\,.
\end{equation}
The corresponding contribution to the spectral form factor
is easily evaluated as
\begin{equation}
\label{structure_contribution} K(\tau)\big|_{\rm
fam}=\frac{(-1)^V\prod_l l^{v_l}}{(L-V-1)!L}\tau^{L-V+1}\,.
\end{equation}
We have thus rederived one of the main results of \cite{EssenFF} in
the elegant fashion suggested by the present work on transport.

\end{document}